Title: Using Deposition Rate and Substrate Temperature to Manipulate Liquid Crystal-like Order in a Vapor-deposited Hexagonal Columnar Glass

Authors: Camille Bishop*,[1] Zhenxuan Chen,[2] Michael F. Toney,[3] Harald Bock,[4] Lian Yu,[1,2] M.D. Ediger[1]

Corresponding Author email: camille.bishop@nist.gov

1. University of Wisconsin – Madison, Department of Chemistry, 1101 University Ave, Madison, WI, 53706 (USA)
2. University of Wisconsin – Madison, School of Pharmacy, 777 Highland Ave, Madison, WI, 53705 (USA)
3. University of Colorado Boulder, Department of Chemical and Biological Engineering, Boulder, CO 80309 (USA)
4. Centre de Recherche Paul Pascal, CNRS & Université de Bordeaux, 115, av. Schweitzer, 33600 Pessac (France)

Abstract

We investigate vapor-deposited glasses of a phenanthroperylene-ester, known to form an equilibrium hexagonal columnar phase, and show that liquid crystal-like order can be manipulated by the choice of deposition rate and substrate temperature during deposition. We find that rate-temperature superposition (RTS) – the equivalence of lowering deposition rate and raising substrate temperature – can be used to predict and control the molecular orientation in vapor-deposited glasses over a wide range of substrate temperatures ($0.75T_g$ to $1.0T_g$). This work extends RTS to a new structural motif, hexagonal columnar liquid crystal order, which is being explored for organic electronics applications. By several metrics, including the apparent average face-to-face nearest-neighbor distance, PVD glasses of the phenanthroperylene-ester are as ordered as the glass prepared by cooling the equilibrium liquid crystal. By other measures, the PVD glasses are less ordered than the cooled liquid crystal. We explain the difference in the maximum attainable order with the existence of a gradient in molecular mobility at the free surface of a liquid crystal, and its impact upon different mechanisms of structural rearrangement. This free surface equilibration mechanism explains the success of the RTS principle and provides guidance regarding the types of order most readily enhanced by vapor deposition. This work extends the applicability of RTS to include molecular systems with a diverse range of higher-order liquid crystalline morphologies that could be useful for new organic electronic applications.

Introduction

As the need for energy-efficient organic electronics grows, materials with new and easily tunable properties are in high demand. One class of materials that has attracted significant interest is highly anisotropic glasses. Anisotropic glasses combine the advantageous properties of crystals such as significant positional correlation[1-2] and preferential orientation[3] with properties of isotropic glasses, such as macroscopic homogeneity[4] and compositional flexibility.[5] This combination of properties makes anisotropic glasses uniquely suited for certain organic electronic applications such as organic light emitting diodes (OLEDs).[6-7] Other devices, such as those based upon organic field effect transistors (OFETs), may benefit from even more highly anisotropic glasses to direct charge mobility.[8-9]

One way to prepare a highly anisotropic glass is by using liquid crystal mesogens. Liquid crystal (LC) glasses are highly anisotropic solids conventionally prepared from a fluid liquid crystal phase[10-11] in which the molecules form mesophases with orientational, and sometimes positional, order. Many liquid crystal morphologies exist, ranging from the simplest orientationally ordered nematic to complex mesophases with both orientational and positional order.[12-14] Columnar liquid crystals are one of the highly organized morphologies that has been gaining increasing interest for use in preparation of glasses for organic electronic applications, specifically OFETs.[8] In these columnar LC glasses, the disc-like molecules assemble face-to-face into columns, and the columns further organize into mesoscale geometric motifs such as rectangular and hexagonal structures. Such glasses can be prepared by cooling from the melt,[15] though it can be difficult to prepare glasses with well-aligned structures through this route.[16] High levels of structural organization are important for applications in devices for directional charge mobility.[8, 17] Traditionally LC systems are aligned by surface contact with an alignment layer, or by application of an electric or magnetic field.[18]

One strategy that has been used to successfully control the order and alignment of highly organized columnar[1, 9] and smectic[19-20] liquid crystal glasses is physical vapor deposition (PVD). PVD can prepare LC glasses with a wide range of orientations, from molecules lying flat on a substrate to end- or edge-on arrangements. Highly organized LC and LC-like glasses with a wide range of orientations and order have been prepared by changing the substrate temperature[1, 19-20] and/or rate[21-22] of deposition. More specifically, glasses with a wide range of rectangular and hexagonal columnar structures have been previously prepared by changing the substrate temperature during deposition.[1] Using PVD to produce highly ordered glasses of mesogens is advantageous because the as-deposited order is independent of the underlying substrate, thereby requiring no surface modification, and PVD can be used to prepare glasses at a processing temperature below their glass transition temperature, saving energy and preserving the structure of the other layers. Additionally, vapor-deposition has been shown to improve and optimize charge mobility in organic electronic columnar systems.[9] The molecular alignment in PVD glasses can be explained by the surface equilibration mechanism, in which enhanced mobility at the free surface of the glass, relative to the bulk,[23-24] allows newly deposited molecules to partially (or fully) equilibrate before becoming trapped in the bulk by further deposition. If the equilibrium free surface is highly anisotropic, this leads to a highly anisotropic as-deposited glass,[25] as predicted by simulation[26-28] and verified by experiment.[29] This results in anisotropic glasses prepared from molecules with[1, 19-20] and without[3, 27] equilibrium liquid crystal phases.

The concept of "Deposition Rate-Substrate Temperature Superposition" (RTS) further extends the utility of the surface equilibration mechanism. If a structural property obeys RTS, decreasing the deposition rate and raising the substrate temperature during deposition will alter the structure of the as-deposited glass in an equivalent manner.[21-22] Increasing the substrate temperature increases the mobility of the newly-deposited surface molecules, decreasing the amount of time required for them to equilibrate. Decreasing the deposition rate allows the newly deposited molecules more time at the free surface to equilibrate, thereby having the same effect. RTS can then be used to predict and control glass structure over a wide range of deposition conditions. When RTS applies, it effectively reduces a two-dimensional parameter space to a single dimension and thus is a major simplification. Previous studies established the existence of RTS for both a mesogenic molecule with equilibrium liquid crystal phases[21] and a non-mesogenic molecule without liquid crystal phases[22] over a rather modest range of substrate temperatures – $0.94T_g$ to $1.00T_g$, where $T_g$ is the glass transition temperature.

Here we show the effect of both substrate temperature and deposition rate on the structure of a vapor-deposited phenanthroperylene-ester that forms a hexagonal columnar liquid crystal glass. We show that both the temperature and rate have a major impact on the molecular orientation, face-to-face nearest-neighbor distance, and hexagonal columnar order in the vapor-deposited glass. We show that the orientational order of vapor-deposited phenanthroperylene-ester obeys the rate-temperature superposition (RTS) principle over a wide range of substrate temperatures– from 0.75 to 1.0 $T_g$. Other structural parameters (apparent nearest-neighbor distance and hexagonal columnar order) are also consistent with RTS and can be precisely tuned with proper choice of deposition conditions. Interestingly, some measures of order for the PVD glasses reach the values for the equilibrium liquid crystal, while others do not. These observations, together with the surface equilibration mechanism, indicate the presence of a gradient in mobility that controls the structural order that may be realized in vapor-deposited glasses. These new results may guide strategies to optimize the organization of columnar glasses for applications in organic electronics.

## Methods

### *Material synthesis.*

A phenanthro[1,2,3,4,*ghi*]perylene-1,6,7,12,13,16-hexacarboxylic hexaester was synthesized as described in Kelber et. al.[15] The melting point of its crystalline solid is $T_m$ = 520 K. The supercooled liquid exhibits a hexagonal columnar phase below 492 K which undergoes a glass transition at $T_g$ = 392 K. For the remainder of this publication, we refer to this molecule as "phenanthroperylene-ester".

### *Vapor Deposition*

Glasses of the phenanthroperylene-ester were prepared in a custom-built deposition chamber as outlined in a previous publication,[1] using a source-to-substrate distance of 11 cm. Samples were deposited on 1-inch Si <1 0 0> wafers with a native oxide (Virginia Semiconductor). The Si wafers were affixed using Apiezon H grease to copper blocks cooled by liquid nitrogen. The temperature of the blocks was controlled using a Lakeshore 336 PID controller. The deposition rate was monitored *in situ* using a Quartz Crystal Microbalance (Sycon Instruments). Average sample thicknesses ranged from 35 nm to 120 nm. The deposition geometry introduced a thickness gradient. The thickness at various points on the wafer was measured by ellipsometry and used to calculate the true deposition rate at each position.

*Ellipsometry*

An M-2000U spectroscopic ellipsometer from J.A. Woollam Co., Inc., equipped with a translational stage, was used to perform measurements on the as-deposited glasses at 21 different positions on each 1-inch wafer. For each measurement, we utilized 7 angles of incidence between 45° and 75°. Measurements were made perpendicular to the thickness gradient across the wafer to minimize beam depolarization effects.

Ellipsometry data were collected over wavelengths from 200 – 1000 nm. Due to a strong absorption at ~350 nm, data was modeled from 500 – 1000 nm using an anisotropic Cauchy model

$$n_z = A_z + \frac{B}{\lambda^2}, n_{xy} = A_{xy} + \frac{B}{\lambda^2}$$

in which $A_z$, $A_{xy}$, and B are empirical optical constants. The difference $A_z$ - $A_{xy}$ is determined in difference mode in CompleteEASE,(J.A. Woollam Co.) and is equal to the birefringence Δn. Films thinner than 90 nm did not provide unique birefringence values, so in Figure 2 we present data from positions on films that are thicker than 90 nm. In Figure 2, the data for a given sample is presented as a single point; the deposition rate is equal to the average of all points on a sample that are thicker than 90 nm and the error bars for the birefringence is the standard deviation of measured values across these spots.

*X-ray scattering*

GIWAXS was performed at Beamline 11-3 of the Stanford Synchrotron Radiation Laboratory (SSRL). The wavelength used is 0.976 Å, with an energy of 12.6 keV. The source to detector distance was 315 mm. A fixed incidence angle of $\theta_{in}$ = 0.12° was used, above the critical angle for phenanthroperylene-ester and below that of the silicon substrate. Scans were performed in a helium atmosphere for exposure times ranging from 30 s to 700 s. Structural parameters presented in this work – the orientational order parameter, nearest-neighbor distance, and hexagonal order – are quantitatively consistent for all exposure times.

Two-dimensional X-ray data was reduced using the WAXSTools plugin[30] in Nika[31-32] run in the Igor Pro environment. The raw area data is processed into reciprocal space by accounting for the scattering geometry.[33]

The orientational order parameter $S_{GIWAXS}$[34] was calculated to quantify the anisotropic scattering of the peak at q ~ 1.8 $Å^{-1}$, which corresponds to the orientation of face-to-face nearest-neighbor interactions. All patterns were integrated from q = 1.65 to 1.85 $Å^{-1}$ for each azimuthal angle $\chi$, with $\chi = 0°$ defined along $q_z$. For background subtraction, similar arcs from q = 1.55 to 1.65 $Å^{-1}$ and q = 1.85 to 1.95 $Å^{-1}$ were integrated and averaged, and then subtracted at each value of $\chi$. The resulting I vs. $\chi$ curves were then fit to empirical functions and extrapolated to the regions with missing intensity, from $\chi = 0°$ to 10° and $\chi = 86°$ to 90°. These procedures are shown in SI Section 1. The resulting functions were used to calculate the sine-corrected $<\cos^2\chi>$ average of the scattering intensity:

$$\langle cos^2\chi\rangle = \frac{\int_0^{90} I(\chi)(cos^2\chi)(sin\chi)d\chi}{\int_0^{90} I(\chi)(sin\chi)d\chi} \quad (1)$$

to determine $S_{GIWAXS}$:

$$S_{GIWAXS} = \frac{1}{2}(3\langle cos^2\chi\rangle - 1) \quad (2)$$

Values in the limit of +1 indicate the face-to-face stacking direction is perpendicular to the substrate, and the limit of -0.5 indicates a stacking direction in-the-plane of the substrate.

To determine the apparent nearest-neighbor distance, we inspected the $q = 1.65$ to $1.85$ Å$^{-1}$ integrations described in the previous paragraph. For each individual sample, we determined the $\chi$ range over which the intensity was at least 50% of the maximum scattering intensity; we limit our analysis to this region in order to report the nearest-neighbor distance for the major structural motif of the sample. We integrated the 2D scattering pattern over this $\chi$ range to yield I vs. q curves. An empirical background subtraction was performed on each I vs. q curve, and the maximum in intensity determined by the first derivatives was used to determine $q_{max}$. $d_{nn}$, the apparent nearest-neighbor distance, was calculated as $d_{nn} = 2\pi/q_{max}$. More information is presented in SI Section 2.

To determine the hexagonal order, the scattering intensity was integrated from $q = 0.35$ to $0.55$ Å$^{-1}$ at all azimuthal angles $\chi$. A constant baseline, equal to the minimum intensity at any scattering angle, was subtracted. Peaks at $\chi \sim 60°$ were fit to Gaussians to extract the full-width at half maximum, denoted in this publication as $\Delta_\chi$.

*Liquid-cooled glass preparation*

Reference liquid-cooled glasses (used to obtain data shown in Figures 2-4) were prepared by annealing a PVD glass in a helium atmosphere at SSRL beamline 11-3, with *in situ* measurements during heating and cooling. Briefly, a sample of phenanthroperylene-ester vapor-deposited at 392 K ($T_g$) and $10^{0.6}$ Å $s^{-1}$ was heated to 457 K to form the equilibrium hexagonal columnar liquid crystal, and then cooled to 387 K at approximately 2.5 K/min (to form the liquid-cooled glass). Quantities shown in Figures 2-4 are measured from the glass at 387 K, 5 K below $T_g$. Below $T_g$, liquid crystalline structural features in glasses are essentially fixed,[11] meaning that the structure probed just below $T_g$ is equivalent that of the equilibrium liquid crystal just above $T_g$.

## Results

### *Overview of Glassy Structures Prepared by PVD*

Vapor-deposition of phenanthroperylene-ester at different substrate temperatures and deposition rates prepares glasses with a wide range of glassy packings, summarized qualitatively in Figure 1. The molecular structure and phase transition temperatures of phenanthroperylene-ester are shown in Figure 1A. Glasses were vapor-deposited at substrate temperatures ($T_{sub}$) from 293 K to 392 K (0.75$T_g$ to 1.0$T_g$) using deposition rates from $10^{-0.3}$ to $10^{0.7}$ Å $s^{-1}$. Two-dimensional grazing incidence wide-angle X-ray scattering (GIWAXS) patterns for a subset of these glasses are shown in Figure 1B, with illustrations of proposed structures for the extreme cases shown in Figure 1C. Changing $T_{sub}$ and the deposition rate during deposition produces a wide range of glassy packings, with lower deposition rates and higher substrate temperatures resulting in more ordered glasses. Given the common view that glasses are isotropic materials, we briefly explain why all the materials in Figure 1B are described here as glasses. When a liquid crystal is cooled, either a 3D crystal or a glass may be formed. For both liquid crystals and isotropic liquids, glass formation occurs upon cooling when molecular motion becomes sufficiently slow (as long as crystal nucleation is avoided) and this transition is accompanied by a significant drop in the heat capacity.[10, 11] While an isotropic liquid yields an isotropic glass upon cooling, the glass formed from a liquid crystal inherits the anisotropic structure of the liquid crystalline phase and different cooling rates can yield glasses with different levels of liquid-crystalline order.[11] For phenanthroperylene-ester, the solids shown in Figure 1B form the hexagonal columnar liquid crystal when heated above $T_g$. In addition, the most ordered glasses produced by PVD are very similar in structure to the glass produced by cooling the equilibrium liquid crystal into the glassy state (see below). Thus, the description of the materials in Figure 1B as glasses is consistent with literature practice: there is no contradiction between “glass” and “liquid-crystalline order”.

By changing the deposition conditions, we can prepare glasses with varying degrees of oriented hexagonal order, as seen in the X-ray scattering patterns. Hexagonal order can be observed in the GIWAXS patterns shown in Figure 1 from the peak at $q \sim 0.4$ Å$^{-1}$ (~ 16 Å in real space, the approximate diameter of the “disc” formed by one molecule). In glasses with oriented hexagonal order, local maxima of the peak occur at $\chi \sim 0°$ and $\chi \sim \pm 60°$. Three out of the six patterns in Figure 1B qualitatively show hexagonal order; those deposited at 380 K at both rates, and the glass deposited at 360 K at the lower rate of $10^{-0.4}$ Å s$^{-1}$. An illustration of the proposed oriented hexagonal order in a glass deposited at high $T_{sub}$ and low rate is shown at the bottom of Figure 1C. Due to the grazing-incidence geometry, we are only sensitive to in-plane hexagonal packing (column direction parallel to substrate).

The substrate temperature and deposition rate can also be changed to prepare glasses with differing orientations of nearest-neighbor interactions. The azimuthal distribution of the broad peak around 1.8 Å$^{-1}$ in the patterns shown in Figure 1B shows the primary direction of face-to-face interactions of the disc-like molecules. This corresponds to a real-space distance of ~3.5 Å, the interaction associated with $\pi$-$\pi$ molecular orbital overlap of the molecules. At low $T_{sub}$ and high deposition rate (315 K and $10^{0.6}$ Å s$^{-1}$), the intensity is highest nearer to $q_z$, suggesting that the discs primarily stack face-to-face out-of-plane, as illustrated at the top of Figure 1C. As the deposition rate is lowered to $10^{-0.5}$ Å s$^{-1}$ at the same $T_{sub}$, the intensity begins to localize along $q_{xy}$, indicating the transition to in-plane face-to-face interactions. As $T_{sub}$ is increased while the deposition rate is held nearly constant, a similar increased localization in-plane occurs, as seen going across rows in Figure 1B. Our work confirms a previous report showing that a wide range of glassy packings can be prepared from phenanthroperylene-ester by changing $T_{sub}$.[1] Here we show, in addition, that we can use deposition rate to prepare different glassy structures at the same $T_{sub}$.

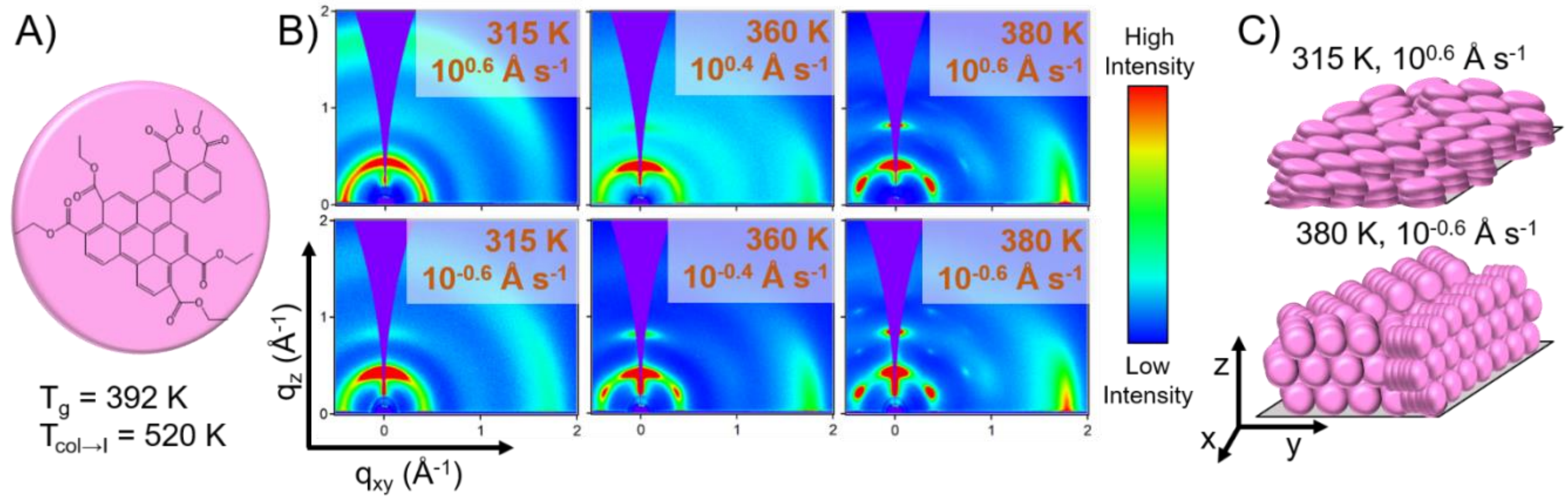

Figure 1. Two-dimensional grazing-incidence wide-angle X-ray scattering (GIWAXS) of vapor-deposited phenanthroperylene-ester shows that both substrate temperature ($T_{sub}$) and deposition rate change the molecular packing, including both orientational and hexagonal order. A) Molecular structure and transition temperatures[1] of phenanthroperylene-ester. B) GIWAXS patterns of phenanthroperylene-ester vapor-deposited at several different substrate temperatures and rates. C) Proposed packing motifs of the lowest $T_{sub}$, highest deposition rate, and the highest $T_{sub}$, lowest deposition rate sample, with real-space coordinate system shown.

*Orientational Order.*

We can quantify the orientation of nearest-neighbor interactions between adjacent phenanthroperylene cores using the orientational order parameter $S_{GIWAXS}$. $S_{GIWAXS}$ is calculated from the anisotropic scattering of the peak at q ~ 1.8 Å$^{-1}$ as defined in *Methods.* Perfectly out-of-plane π-π stacking would correspond to $S_{GIWAXS}$ = 1, while perfect in-plane stacking would produce $S_{GIWAXS}$ = -0.5. $S_{GIWAXS}$ for phenanthroperylene-ester vapor-deposited at many $T_{sub}$ and deposition rates are presented in Figure 2A. As the deposition rate is lowered and $T_{sub}$ is increased, $S_{GIWAXS}$ values show that π-π stacking increasingly occurs in-plane. For example, at $T_{sub}$ = 315 K, $S_{GIWAXS}$ is positive at the highest deposition rate of $10^{0.7}$ Å $s^{-1}$, indicating out-of-plane π-π stacking. As the deposition rate is lowered to $10^{-0.6}$ Å $s^{-1}$, $S_{GIWAXS}$ reaches the negative value of -0.2, with stacking that is primarily in-plane. At constant deposition rate, glasses prepared at higher $T_{sub}$ have more negative values of $S_{GIWAXS}$, consistent with previous reports.[1] At the highest $T_{sub}$ and lowest deposition rate, $S_{GIWAXS}$ reaches -0.45, which is very close to the limiting value of -0.5 associated with perfect in-plane π-π stacking. This value of $S_{GIWAXS}$ is the same as that for the liquid-cooled glass, indicating that vapor deposition at low rates and high $T_{sub}$ results in a glass with orientation similar to that prepared by cooling the equilibrium liquid crystal.

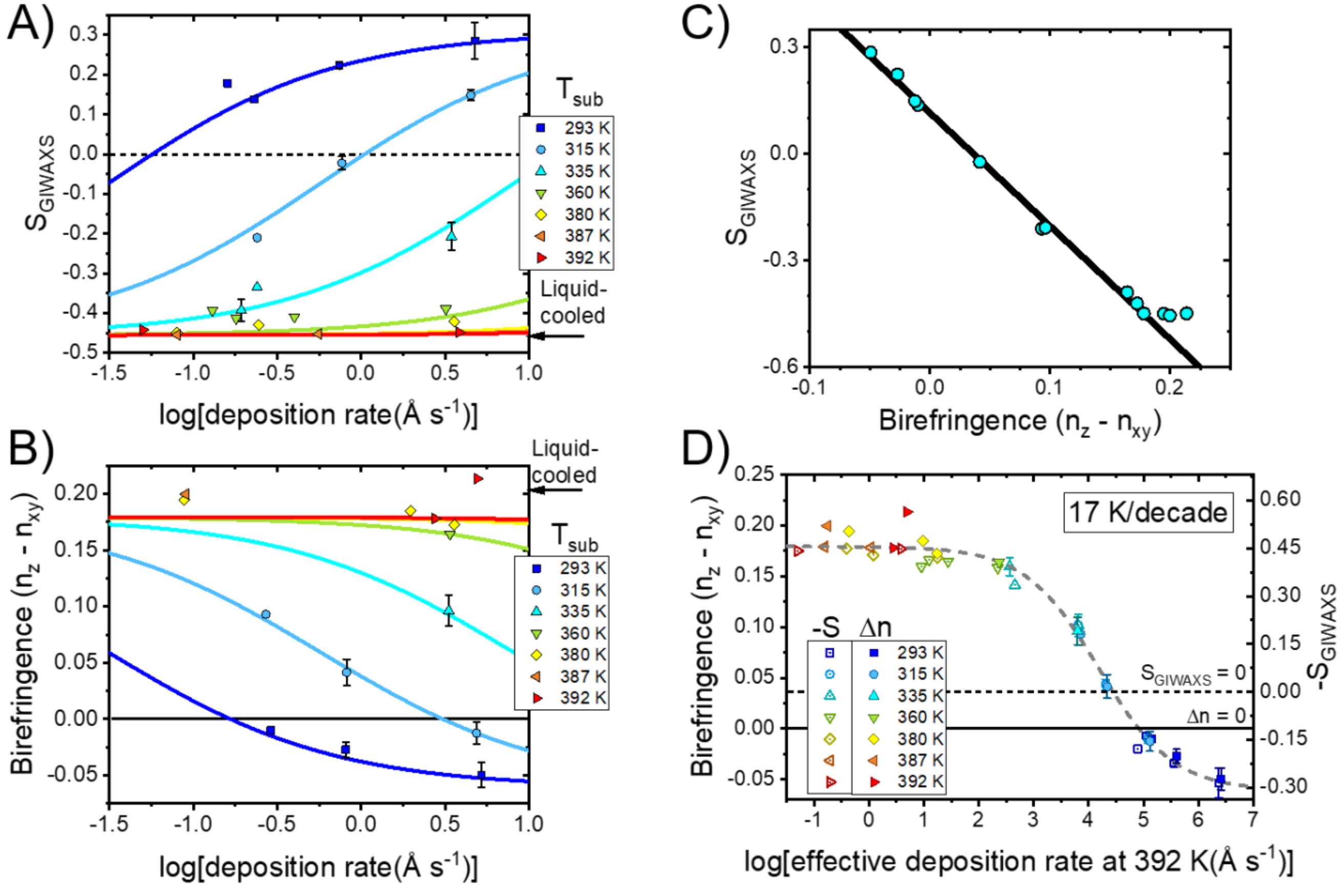

Figure 2. Measures of orientational order of vapor-deposited phenanthroperylene-ester depend on both $T_{sub}$ and deposition rate, and show deposition rate-substrate temperature superposition (RTS). A) The GIWAXS-derived orientational order parameter $S_{GIWAXS}$ for the q ~ 1.8 $Å^{-1}$ (face-to-face nearest-neighbor interactions) peak vs. the log of the deposition rate at several substrate temperatures and for the liquid-cooled glass. $S_{GIWAXS}$ ranges from 0.3 at low $T_{sub}$ and high rate, indicating mildly out-of-plane π-π stacking, to -0.45 at high $T_{sub}$ and low rate, consistent with nearly perfect in-plane π–π stacking (along $q_{xy}$). B) Optical birefringence vs. the log of the deposition rate at several substrate temperatures and for the liquid-cooled glass. A more positive birefringence, observed at high $T_{sub}$ and low rate, indicates a higher refractive index out-of-plane

than in-plane, consistent with edge-on molecular orientation. Error bars presented are the standard deviation of the birefringence across the wafer. For points where error bars are not shown, the error bars are smaller than the size of the symbol. C) $S_{GIWAXS}$ vs. birefringence for all the data shown in panels A & B. A linear fit is applied to all data except for that with $S_{GIWAXS} < -0.44$, due to uncertainty of calculating $S_{GIWAXS}$ for highly oriented samples. D) Combined plot of $S_{GIWAXS}$ and birefringence values vs. the log of the effective deposition rate at 392 K ($T_g$). The effective deposition rate for both parameters is calculated using a shift factor of 17 K per decade. An empirical hyperbolic tangent fit is shown as a guide to the eye and is also shown as colored lines in panels A and B.

The optical birefringence, shown in Figure 2B, provides a complementary measure of orientational order that reinforces the X-ray scattering. The birefringence is sensitive to molecular orientation because of the polarizability of the phenanthroperylene core is different in the in- and out-of-plane directions. Previous studies[3, 19] have shown that the birefringence can have a one-to-one correlation with the orientation of the long axis of the molecule. Similar to the data shown in Figure 2A, the birefringence switches sign as the deposition conditions change from low $T_{sub}$ and high deposition rate to high $T_{sub}$ and low deposition rate. At high $T_{sub}$/low deposition rates, the birefringence is positive (refractive index is greater out-of-plane than in-plane), indicating edge-on molecular orientation. Since the plane of the molecule is perpendicular to the nearest-neighbor interactions, this is consistent with the negative $S_{GIWAXS}$ parameter indicating that the interactions are primarily in-plane. We note that, in Figure 2B, the observed birefringence values reach the value of the liquid-cooled glass (the lines through the data show slightly different behavior, as explained below).

The strong correlation between $S_{GIWAXS}$ and the birefringence is shown in Figure 2C. The two variables are almost perfectly anti-correlated, as expected given that they are both sensitive to the same structural feature. We note that the lowest $S_{GIWAXS}$ points do not fit on this line. Because the molecules in these samples have stacking so strongly localized in-plane, with the scattering concentrated at $\chi > 86°$, we cannot precisely determine the value of $S_{GIWAXS}$. We expect that the true values of $S_{GIWAXS}$ saturate at -0.5 (the minimum possible value) at the highest $T_{sub}$ and lowest deposition rates, even though the experimentally-determined $S_{GIWAXS}$ values plateau at -0.45.

We find that a rate-temperature superposition (RTS) successfully relates the effects of $T_{sub}$ and deposition rate on orientational order in the as-deposited phenanthroperylene-ester glasses, as shown in Figure 2D. Using the anti-correlation found in Figure 2C, we plot the $S_{GIWAXS}$ and birefringence data on linked axes. We find that the data presented in Figures 2A and 2B can be collapsed into a single master curve with a shift factor of 17 K/decade, i.e., by equating the orientational order prepared by a 17 K rise in $T_{sub}$ to that resulting from depositing 10 times more slowly. This allows the calculation of an "effective deposition rate" at 392 K, which is the deposition rate that would be required to prepare a glass with the given structure at $T_{sub}$ = 392 K. The temperature dependence of the relaxation time suggested by this shift factor is much weaker than that found for previous liquid crystalline systems, consistent with a highly mobile surface, as elaborated further in the Discussion section.

One interesting feature of the data shown in Figure 2D is that the zero for birefringence and $S_{GIWAXS}$ do not occur at the same effective deposition rate. Where $S_{GIWAXS}$ is equal to zero, consistent with azimuthally isotropic scattering from the q ~ 1.8 Å$^{-1}$ feature, the birefringence $n_z - n_{xy}$ is equal to 0.04. This may seem counter-intuitive, since a truly isotropic sample would have both $S_{GIWAXS}$ and the birefringence equal to zero. The resolution to this conundrum is the observation that none of the vapor-deposited glasses are truly isotropic. For example, SI Figure S7 shows the 2D scattering pattern for the glass deposited at 315 K and $10^{-0.1}$ Å $s^{-1}$. Despite the isotropic scattering of the q ~ 1.8 Å$^{-1}$ peak for a glass that is characterized by $S_{GIWAXS} = 0$, other aspects of the GIWAXS pattern are clearly anisotropic. It is possible that not all regions of the sample contribute equally to the scattering, for example, less well-ordered regions of the sample might contribute to the birefringence but not the scattering. Other workers have reported that mildly birefringent glasses can be prepared with isotropic molecular orientation,[35] consistent with the idea that there are many dimensions to anisotropy in vapor-deposited glasses.

*Average face-to-face nearest-neighbor distance.*

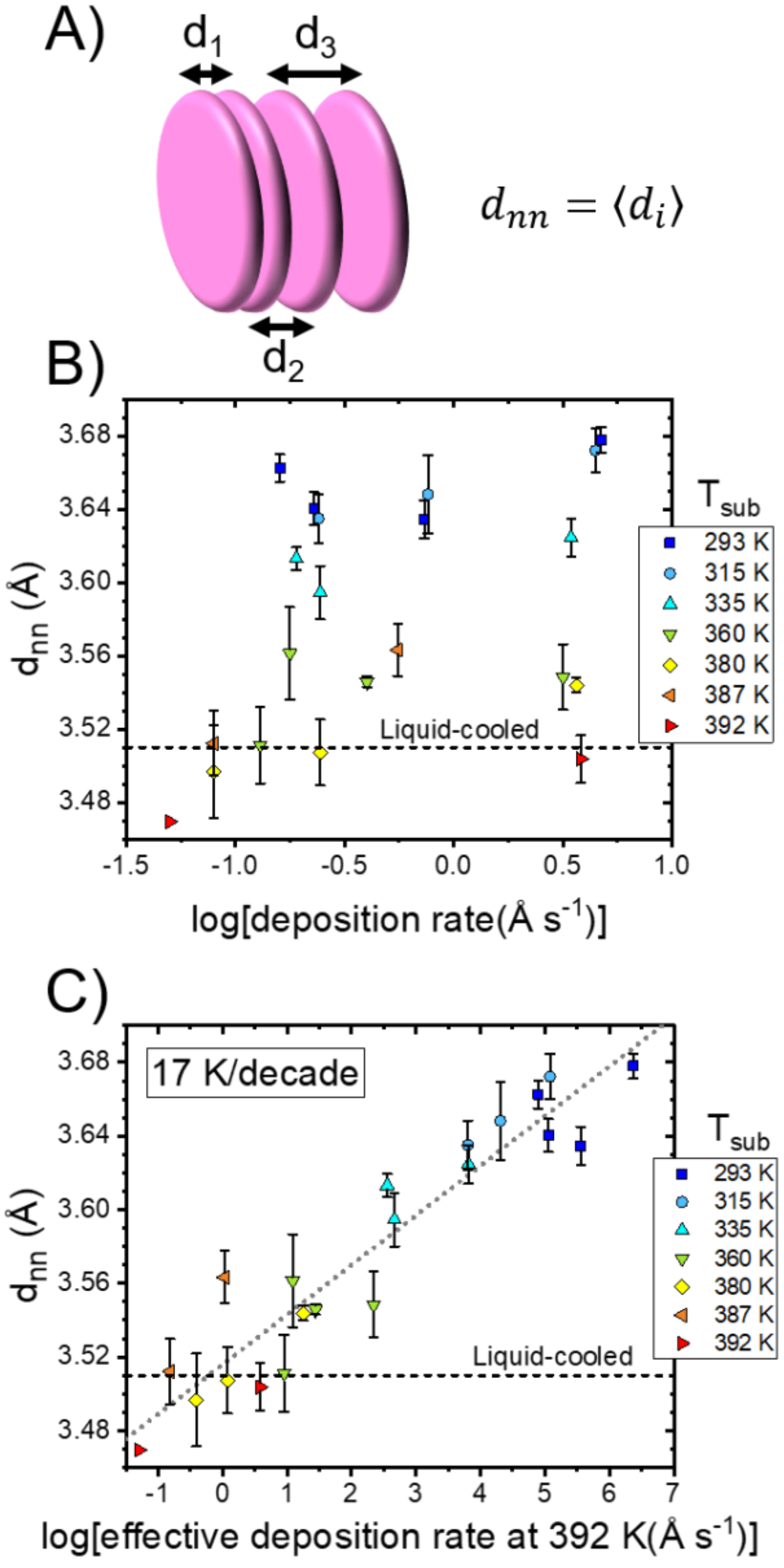

Figure 3. The apparent average nearest-neighbor distance $d_{nn}$ is significantly modified by deposition rate and substrate temperature and can be described using deposition rate-substrate temperature superposition (RTS). A) Illustration quantifying $d_{nn}$ – the average distance between two adjacent disc-like molecules in the vapor-deposited glass. B) $d_{nn}$ of vapor-deposited phenanthroperylene-ester as a function of deposition rate at several $T_{sub}$. Generally, $d_{nn}$ decreases at higher $T_{sub}$ and lower deposition rates. The value of $d_{nn}$ for the liquid-cooled glass is also indicated. Error bars show the standard deviation of the peak position found from the measurements at the distinct spots on the sample. C) RTS successfully describes $d_{nn}$ using a shift factor of 17 K/decade, the same shift factor used for molecular orientation in Figure 2. A linear fit to all data points is shown as a guide to the eye.

Both increasing the substrate temperature and decreasing the deposition rate decrease the face-to-face nearest-neighbor distance $d_{nn}$ in vapor-deposited glasses of the phenanthroperylene-ester. This apparent distance is defined from the q-value at which the maximum scattering occurs, where $d_{nn} = 2\pi/q_{max}$. We expect that the smallest $d_{nn}$ values (which occur in the most highly ordered samples) represent the π-stacking distance. In this regime, a smaller $d_{nn}$ can result in an increase in π-orbital overlap between adjacent molecules, enhancing charge mobility and performance.[36] Figure 3A shows a schematic illustration of $d_{nn}$ in vapor-deposited phenanthroperylene-ester, while $d_{nn}$ is plotted against the log of the deposition rate in Figure 3B. At a constant deposition rate of $\sim 10^{0.5}$ Å $s^{-1}$, glasses deposited at the lowest $T_{sub}$ of 293 K have $d_{nn}$ equal to 3.68 Å, while those deposited at the highest $T_{sub}$ of 392 K have a $d_{nn}$ of 3.51 Å – a 4.5% decrease. For all $T_{sub}$ above 293 K, a decrease in deposition rate generally results in a decrease in $d_{nn}$. The values of $d_{nn}$ approach and, at the highest substrate temperature and lowest deposition rate, are even lower than the value of the liquid-cooled glass.

Using the orientational order shift factor determined in Figure 2, we test the rate-temperature superposition (RTS) for $d_{nn}$ and find that it works reasonably well, as shown in Figure 3A. The deposition rate of the glasses has been converted to an effective deposition rate at 392 K using a shift factor of 17 K/decades. Given the relatively large error bars for $d_{nn}$, we did not attempt to fit the shift factor. Thus, our analysis indicates that the $d_{nn}$ values are consistent with shift factor found for molecular orientation.

The correlation length associated with the nearest-neighbor peak depends upon deposition conditions in a manner that is qualitatively consistent with data in Figures 2 and 3. We find that the correlation length ranges from about 1.5 nm (~4 molecules) for the sample deposited at the lowest $T_{sub}$ and highest rate, up to about 3.3 nm (~9 molecules), for the sample deposited at the highest $T_{sub}$ and lowest rate.

*Hexagonal order.*

Similar to the molecular orientation and the apparent nearest-neighbor distance, we find that glasses with a higher degree of in-plane hexagonal order are obtained by decreasing the deposition rate and increasing $T_{sub}$. To quantify the hexagonal order, we integrate the peak at $q \sim 0.4$ Å$^{-1}$ (from $q = 0.35$ to $0.55$ Å$^{-1}$) for every azimuthal angle $\chi$, as illustrated schematically in Figure 4A. Example integrations are shown in Figure 4B. There is no intensity data between $\chi = \pm 8°$ due to the scattering geometry,[33] though all samples show a peak near this area regardless of the extent of hexagonal order. As $T_{sub}$ is increased and the deposition rate is decreased, peaks at $\chi \sim 60°$ grow in size and narrow in width. We subtract a linear background and fit these peaks to a Gaussian function, using the full-width half max, denoted as $\Delta_\chi$, as a measure of hexagonal order. Figure 4C shows $\Delta_\chi$ values as a function of the log of deposition rate at several $T_{sub}$. Glasses deposited below 360 K do not show any in-plane hexagonal order and are not included in Figure 4C. At all $T_{sub}$ studied, depositing more slowly prepares glasses with greater in-plane hexagonal order (smaller $\Delta_\chi$). Generally, raising $T_{sub}$ also increases the hexagonal order. All glasses prepared by vapor deposition have less in-plane hexagonal order than the glass prepared by liquid-cooling, with a smallest value $\Delta_\chi$ being 3° as compared to the liquid-cooled value of 1°. Due to the grazing incidence geometry, the intensity of the $\chi \sim 60°$ peaks are sensitive to both the degree of hexagonal order within a domain and the degree to which the hexagonal columns lie in the substrate plane. Additionally, we find that the sample deposited at the highest effective deposition rate has a correlation length of 5.1 nm, or 3 to 4 molecular diameters. The sample deposited at the lowest effective rate has a correlation length of 33 nm, or ~ 20 molecular diameters, suggesting a highly organized sample. We also calculate the edge-to-edge disc spacing of this hexagonal peak around q ~ 0.4 Å$^{-1}$, shown in SI Section 5.

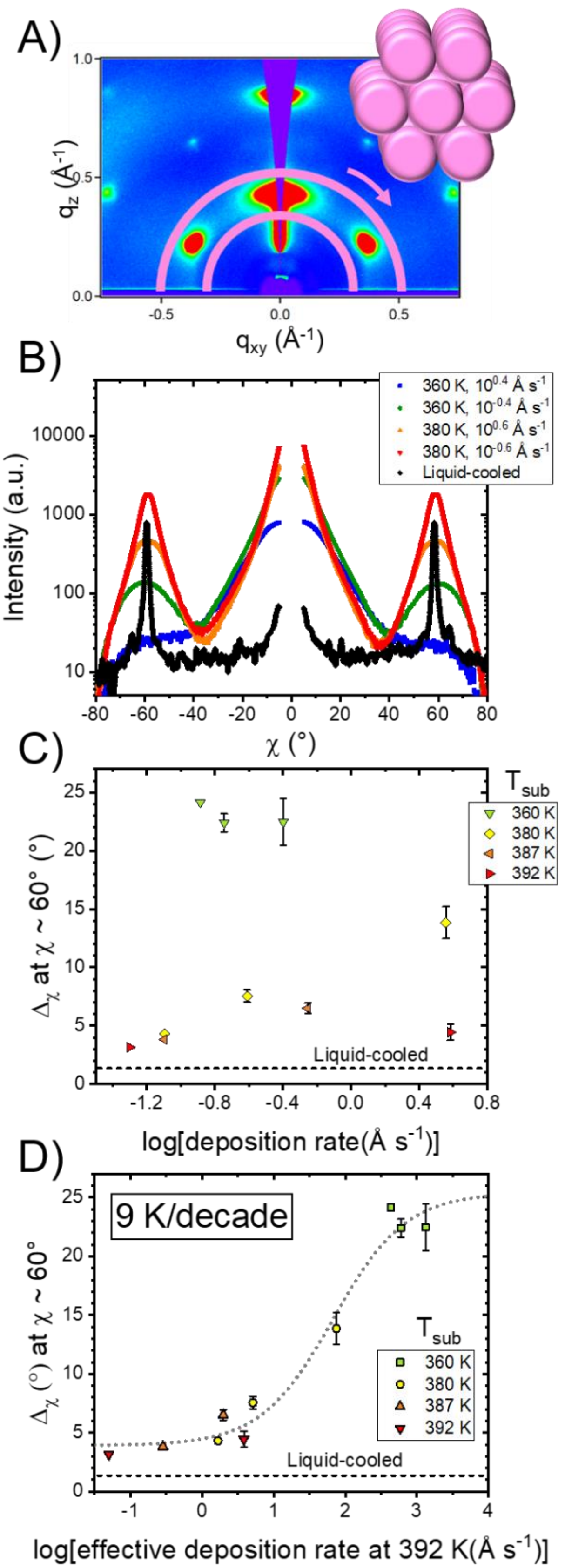
A)
q_z (Å^-1)
q_xy (Å^-1)
B)
360 K, 10^0.4 Å s^-1
360 K, 10^-0.4 Å s^-1
380 K, 10^0.6 Å s^-1
380 K, 10^-0.6 Å s^-1
Liquid-cooled
Intensity (a.u.)
χ (°)
C)
T_sub
360 K
380 K
387 K
392 K
Δ_χ at χ ~ 60° (°)
Liquid-cooled
log[deposition rate(Å s^-1)]
D)
9 K/decade
T_sub
360 K
380 K
387 K
392 K
Δ_χ (°) at χ ~ 60°
Liquid-cooled
log[effective deposition rate at 392 K(Å s^-1)]

Figure 4. Azimuthal integrations of the q ~ 0.4 Å$^{-1}$ peak reveal that in-plane hexagonal order depends upon $T_{sub}$ and deposition rate, and can be described by RTS. A) Illustration of area in the 2D GIWAXS pattern used to determine hexagonal order, and a schematic of a highly order system. Intensity is integrated from q = 0.35 to 0.55 Å$^{-1}$ at every value of the azimuthal angle $\chi$. B) Scattering intensity for selected samples as a function of $\chi$, for vapor-deposited and liquid-cooled glasses. C) FWHM ($\Delta_\chi$) of the peak at $\chi \sim 60°$ as a function of deposition rate for all samples showing hexagonal order. Generally, raising $T_{sub}$ and lowering the deposition rate result in a smaller $\Delta_\chi$, consistent with increasingly perfected in-plane hexagonal order. D) Hexagonal order can be described using RTS. $\Delta_\chi$, the FWHM of the hexagonal order peak at $\chi \sim 60°$, is plotted against the log of the effective deposition rate at 392 K calculated using a shift factor of 9 K per decade in rate. An empirical hyperbolic tangent fit is shown as a guide to the eye.

We test the RTS principle for hexagonal order, and find that $\Delta_\chi$ superposes successfully, but with a different shift factor than that found for the orientational order and nearest-neighbor distance. In Figure 4D, we show $\Delta_\chi$ as a function of the effective deposition rate calculated from a superposition factor of 9 K per decade. While the existence of multiple shift factors may seem surprising, this was previously reported for orientational and positional order in a vapor-deposited smectic liquid crystal,[21] and attributed to the existence of multiple relaxation processes at the free surface. Below we show that these distinct shift factors provide insight into the surface equilibration process during PVD.

## Discussion

*Deposition Rate-Substrate Temperature Superposition of Orientational Order*

This work expands the range of systems for which RTS has been shown to describe orientational order in vapor-deposited glasses. The effect of deposition rate and $T_{sub}$ on the structure of three vapor-deposited systems have now been systematically studied: the rod-like molecules itraconazole (a smectic liquid crystal) and posaconazole (with no liquid crystal phases), and now the disc-like phenanthroperylene-ester, a hexagonal columnar liquid crystal. In contrast to previous work, phenanthroperylene-ester is a potential organic photovoltaic material with favorable electronic properties.[15] For all three of these systems, the molecules in the vapor-deposited glass orient edge-on with respect to the substrate when prepared at higher $T_{sub}$ and/or lower deposition rate. As these conditions maximize equilibration at the free surface during deposition, this result is consistent with the surface equilibration mechanism for PVD glasses (since edge-on orientation is the typical equilibrium liquid crystal orientation at the free surface).[18]

This work also expands the range of substrate temperature over which RTS for orientational order can be expected to work ($0.75T_g$ to $1.00T_g$). Previous work by Bishop et. al. covered a more modest range of substrate temperatures from $0.94T_g$ to $1.00T_g$.[21-22] Previous work by Yokoyama and coworkers, while not explicitly investigating RTS, found different birefringence values for OLED materials TPT1, 2-TNATA, and TPD deposited at different rates at a single substrate temperature ($0.70T_g$, $0.77T_g$, and $0.88T_g$, respectively).[6, 37] Additionally, Bagchi et. al. found differing *translational* order at two deposition rates at several substrate temperatures for $Alq_3$, a spheroidal OLED molecule.[27] The results from these works on OLED molecules together with the present work suggest that RTS may be applied systematically to technologically relevant materials over a wide range of deposition conditions.

*Comparison of PVD Glasses with Glasses Prepared by Cooling the Equilibrium Liquid Crystal*

To evaluate the structures of PVD glasses of the phenanthroperylene-ester, it is convenient to use the equilibrium liquid crystal as a reference state representing perfect liquid crystal order. For two of the types of structural order characterized in this work, PVD glasses can achieve the same level of structural perfection as glasses prepared by cooling the equilibrium liquid crystal. For example, glasses deposited at sufficiently high $T_{sub}$ and sufficiently low rate reach the values for the apparent nearest-neighbor distance, $d_{nn}$, and the orientational order that result from cooling the equilibrium liquid crystal, as measured by $S_{GIWAXS}$ and birefringence. The PVD glass deposited at the lowest rate at 392 K exhibited values of $d_{nn}$ that are even smaller than that found for the equilibrium liquid crystal; this glass presumably is more equilibrated than our reference sample, which has a structure that resembles the liquid crystal at a temperature around 392 K, where molecular motions arrest upon cooling. However, even at the lowest effective deposition rate studied, the hexagonal order as measured by $\Delta_{\chi}$ never reaches the equilibrium value of 1.5°, instead reaching a minimum value of 3° by vapor deposition. The behavior of the different structural parameters can be understood using the surface equilibration mechanism, as we discuss below.

*Surface Equilibration Mechanism and a Gradient in Mobility.*

Organic glasses can have highly mobile surfaces which allow partial or full equilibration during PVD even when the substrate temperature is below $T_g$.[38-40] By this means, glasses with greater density and much higher kinetic stability can be prepared.[23] Since equilibration during PVD occurs near an interface, anisotropic glasses are often obtained. For glasses of phenanthroperylene-ester prepared at high $T_{sub}$ and low deposition rate, conditions that favor equilibration, strongly edge-on packing relative to the free surface is observed. We interpret this to mean that edge-on packing is preferred at the free surface of the equilibrium liquid crystal, and this is consistent with experimental results for other disc-like liquid crystals.[41-42]

The surface equilibration mechanism can be used to rationalize why PVD can more readily perfect certain types of liquid-crystalline order than others. Computer simulations and experiments indicate that there is a strong gradient in mobility near the free surface of an organic glass. Dynamics are most accelerated at the free surface and somewhat slower for a few molecular diameters below the interface, with bulk-like dynamics [43-45] being achieved at greater distance (often thought to be ~ 5 nm).[24] We speculate that the measures of order which fully attain equilibrium values during vapor deposition of the phenanthroperylene-ester, orientation and nearest-neighbor distance, require rearrangement only in the top monolayer (1.5 nm) where mobility is greatest. We imagine that hexagonal order, which requires registry across at least two molecular layers (3 nm), does not reach equilibrium for the same deposition conditions because rearrangement would be required deeper into the film where the dynamics are slower. We note that the region of high mobility (~ 5 nm) near the free surface is much smaller than the coherence length for hexagonal packing (up to 30 nm), and so we infer that it is possible to prepare a thick region of ordered hexagonal structure by equilibrating only a few layers at a time during deposition.

The shift factor for hexagonal order is smaller than that observed for the nearest-neighbor distance and the orientational order, and this supports the idea that hexagonal order requires equilibration to a greater depth during deposition. The gradient in mobility near the surface of an organic glass is associated with a gradient in activation energies. Depending upon the system and the temperature, activation barriers can be two to five times smaller at the surface than in the bulk.[38, 45] We can relate the temperature dependence of the shift factors to this gradient in activation energies. For the orientational ordering of phenanthroperylene-ester, a shift factor of 17 K/decade indicates an $E_a$ of 110 kJ/mol, while the 9 K/decade shift factor for the hexagonal ordering indicates an $E_a$ of 210 kJ/mol. This greater activation energy for the hexagonal ordering suggests that the rearrangement occurs further into the bulk of the film, where barriers to rearrangement are higher.[44]

The gradient in activation energies near the free surface also allows us to understand results in the literature for PVD glasses of itraconazole, a smectic liquid crystal. Bishop et. al. found different shift factors for the orientational and translational order in itraconazole.[21] The shift factor for the orientational order yields an activation energy of 390 kJ/mol, while that for the presumably more long-range process of smectic layering is 920 kJ/mol. In comparison with itraconazole, the substantially lower activation energies reported here for phenanthroperylene-ester are consistent with a highly mobile free surface and indicate that deposition rate is a much more potent factor for controlling structural order.

*Versatility of PVD for Preparing Highly Anisotropic Glasses.*

PVD of phenanthroperylene-ester provides an efficient route to highly anisotropic glasses with desirable properties. In comparison to slow cooling from the equilibrium liquid crystal, PVD offers advantages of low temperature processing and uniform orientation. For example, a glass can be prepared with the equilibrium value for $d_{nn}$ at $T_{sub}$ = 360 K by depositing sufficiently slowly. As equilibrium processing requires transformation in the melt (well above 392 K), this enables a much lower processing temperature to be used. This is very advantageous in a multi-layer manufacturing process in which high temperatures may be deleterious to the other components in a device. In addition, PVD can prepare films with uniform orientation, even for films that are too thick to be aligned using alignment layers in confinement.[18] Even if PVD at a given $T_{sub}$/deposition rate condition doesn't lead to the desired equilibrium property, the as-deposited glass may have sufficient alignment of the molecules to achieve the desired orientation with limited subsequent annealing near $T_g$; if the same glass were prepared by spin-coating or heating a powder, followed by cooling from the melt, it would likely take much longer to reach the desired orientation, if it can be achieved at all.

An additional advantage of preparing glasses by PVD is access to a large range of structures that are likely not accessible by any other practical route. For phenanthroperylene-ester, a wide range of glassy structures are possible, based upon the four structural metrics presented here. We find a wide range of molecular orientations, with nearest-neighbor stacking varying from mildly out-of-plane to strongly in-plane. The nearest-neighbor distance varies from 3.47 to 3.70 Å and likely provides an opportunity to tune charge mobility.[36] $\Delta_\chi$, the measure of hexagonal order in this work, also varies widely from 3° to 24°. It is possible that glasses with some of these out-of-equilibrium properties could also be prepared by fast quenching from the melt, such as has been performed for smectic liquid crystals.[11] However, the required cooling rates are likely incompatible with other requirements in the context of device manufacture.

Additional experimental techniques will be needed to characterize the domain structure of the columns of phenanthroperylene-ester in PVD glasses, including their length. Due to the deposition geometry, columns propagate in all in-plane directions.[1] AFM could be used in conjunction with high resolution electron microscopy[46] to map domain morphology of samples. One might expect that domains would be longer in samples that are deposited more slowly and at higher substrate temperatures; these longer domains would be expected to transport charge more effectively.

Conclusions

Over a wide range of deposition conditions, the glassy packing of vapor-deposited phenanthroperylene-ester can be controlled by changing the deposition rate and the substrate temperature. Deposition at higher substrate temperature and lower rate results in glasses with more edge-on packing, a smaller face-to-face nearest-neighbor distance (likely resulting in greater π-π overlap), and more hexagonal columnar order. These types of order can be understood and controlled using the concept of "rate-temperature superposition", or RTS. The success of RTS can be explained by the surface equilibration mechanism, in which structure in the vapor-deposited glass is a consequence of partial equilibration at the free surface during deposition; the same equilibration can occur either as a result of increasing the substrate temperature or by decreasing the deposition rate. A gradient in mobility from the free surface explains why the different types of order investigated in this work can be perfected to various degrees by vapor deposition. This is the first time the concept of RTS, found initially in model smectic systems, has been extended to a columnar liquid crystal morphology which shows promising electrical properties.[15] For the phenanthroperylene-ester investigated here, RTS works over a large range of substrate temperatures (from $T_g$ down to $0.75T_g$). We therefore show how substrate temperature and deposition rate can be selected to optimize materials for new organic electronic applications.

## Acknowledgements

We gratefully acknowledge LG Chem for support of this research. Additional support was received from NSF through the University of Wisconsin Materials Research Science and Engineering Center (Grant DMR-1720415), and through Grant DMR-1904601. Use of the Stanford Synchrotron Radiation Lightsource, SLAC National Accelerator Laboratory, is supported by the US Department of Energy, Office of Science, Office of Basic Energy Sciences under Contract DE-AC02-76SF00515.

## Supporting Information Description

The Supporting Information is available free of charge at [ ].

- $S_{GIWAXS}$ calculation and background subtraction, determination of nearest-neighbor distances, comparison of X-ray scattering and optical anisotropy, edge-to-edge disc spacing determination. (PDF)

## References

1. Gujral, A.; Gómez, J.; Ruan, S.; Toney, M. F.; Bock, H.; Yu, L.; Ediger, M. D., Vapor-Deposited Glasses with Long-Range Columnar Liquid Crystalline Order. *Chemistry of Materials* **2017,** *29* (21), 9110-9119.
2. Eccher, J.; Faria, G. C.; Bock, H.; Von Seggern, H.; Bechtold, I. H., Order Induced Charge Carrier Mobility Enhancement in Columnar Liquid Crystal Diodes. *ACS Applied Materials & Interfaces* **2013,** *5*, 11935-11943.
3. Dalal, S. S.; Walters, D. M.; Lyubimov, I.; de Pablo, J. J.; Ediger, M. D., Tunable molecular orientation and elevated thermal stability of vapor-deposited organic semiconductors. *Proceedings of the National Academy of Sciences* **2015,** *112* (14), 4227.
4. Bolognesi, A.; Berliocchi, M.; Manenti, M.; Carlo, A. D.; Lugli, P.; Lmimouni, K.; Dufour, C., Effects of grain boundaries, field-dependent mobility, and interface trap states on the electrical characteristics of pentacene TFT. *IEEE Transactions on Electron Devices* **2004,** *51* (12), 1997-2003.
5. Schmidt, T. D.; Lampe, T.; Sylvinson M. R, D.; Djurovich, P. I.; Thompson, M. E.; Brütting, W., Emitter Orientation as a Key Parameter in Organic Light-Emitting Diodes. *Physical Review Applied* **2017,** *8* (3), 037001.
6. Yokoyama, D., Molecular orientation in small-molecule organic light-emitting diodes. *Journal of Materials Chemistry* **2011,** *21*, 19187-19202.
7. Ràfols-Ribé, J.; Will, P.-A.; Hänisch, C.; Gonzalez-Silveira, M.; Lenk, S.; Rodríguez-Viejo, J.; Reineke, S., High-performance organic light-emitting diodes comprising ultrastable glass layers. *Science Advances* **2018,** *4* (5), eaar8332.
8. O'Neill, M.; Kelly, S. M., Ordered materials for organic electronics and photonics. *Advanced Materials* **2011,** *23*, 566-584.
9. Eccher, J.; Zajaczkowski, W.; Faria, G. C.; Bock, H.; von Seggern, H.; Pisula, W.; Bechtold, I. H., Thermal Evaporation versus Spin-Coating: Electrical Performance in Columnar Liquid Crystal OLEDs. *ACS Applied Materials & Interfaces* **2015,** *7*, 16374-16381.
10. Suga, H.; Seki, S., Thermodynamic investigation on glassy states of pure simple compounds. *Journal of Non-Crystalline Solids* **1974,** *16* (2), 171-194.
11. Teerakapibal, R.; Huang, C.; Gujral, A.; Ediger, M. D.; Yu, L., Organic Glasses with Tunable Liquid-Crystalline Order. *Physical Review Letters* **2018,** *120* (5), 055502.
12. Sorai, M.; Seki, S., Glassy Liquid Crystal of the Nematic Phase of N-(o-Hydroxy-p-methoxybenzylidene)-p-butylaniline. *Bulletin of the Chemical Society of Japan* **1971,** *44* (10), 2887-2887.
13. Tsuji, K.; Sorai, M.; Seki, S., New Finding of Glassy Liquid Crystal – a Non-equilibrium State of Cholesteryl Hydrogen Phthalate. *Bulletin of the Chemical Society of Japan* **1971,** *44* (5), 1452-1452.
14. Farias, G.; Simeão, D. S.; Moreira, T. S.; dos Santos, P. L.; Bentaleb, A.; Girotto, E.; Monkman, A. P.; Eccher, J.; Durola, F.; Bock, H.; de Souza, B.; Bechtold, I. H., An unusual plank-shaped nematogen with a graphene nanoribbon core. *Journal of Materials Chemistry C* **2019,** *7* (39), 12080-12085.
15. Kelber, J.; Achard, M.-F.; Durola, F.; Bock, H., Distorted Arene Core Allows Room-Temperature Columnar Liquid-Crystal Glass with Minimal Side Chains. *Angewandte Chemie International Edition* **2012,** *51* (21), 5200-5203.
16. Al-Lawati, Z. H.; Bushby, R. J.; Evans, S. D., Alignment of a Columnar Hexagonal Discotic Liquid Crystal on Self-Assembled Monolayers. *The Journal of Physical Chemistry C* **2013,** *117* (15), 7533-7539.
17. Bushby, R. J.; Lozman, O. R., Discotic liquid crystals 25 years on. *Current Opinion in Colloid & Interface Science* **2002,** *7* (5), 343-354.
18. Jerome, B., Surface effects and anchoring in liquid crystals. *Reports on Progress in Physics* **1991,** *54* (3), 391-451.
19. Gómez, J.; Jiang, J.; Gujral, A.; Huang, C.; Yu, L.; Ediger, M. D., Vapor deposition of a smectic liquid crystal: highly anisotropic, homogeneous glasses with tunable molecular orientation. *Soft Matter* **2016,** *12*, 2942-2947.
20. Gujral, A.; Gómez, J.; Jiang, J.; Huang, C.; O'Hara, K. A.; Toney, M. F.; Chabinyc, M. L.; Yu, L.; Ediger, M. D., Highly Organized Smectic-like Packing in Vapor-Deposited Glasses of a Liquid Crystal. *Chemistry of Materials* **2017,** *29*, 849-858.

21. Bishop, C.; Gujral, A.; Toney, M. F.; Yu, L.; Ediger, M. D., Vapor-Deposited Glass Structure Determined by Deposition Rate–Substrate Temperature Superposition Principle. *The Journal of Physical Chemistry Letters* **2019,** *10* (13), 3536-3542.
22. Bishop, C.; Li, Y.; Toney, M. F.; Yu, L.; Ediger, M. D., Molecular Orientation for Vapor-Deposited Organic Glasses Follows Rate-Temperature Superposition: The Case of Posaconazole. *The Journal of Physical Chemistry B* **2020,** *124* (12), 2505-2513.
23. Swallen, S. F.; Kearns, K. L.; Mapes, M. K.; Kim, Y. S.; McMahon, R. J.; Ediger, M. D.; Wu, T.; Yu, L.; Satija, S., Organic Glasses with Exceptional Thermodynamic and Kinetic Stability. *Science* **2007,** *315* (5810), 353.
24. Chen, Y.; Chen, Z.; Tylinski, M.; Ediger, M. D.; Yu, L., Effect of molecular size and hydrogen bonding on three surface-facilitated processes in molecular glasses: Surface diffusion, surface crystal growth, and formation of stable glasses by vapor deposition. *The Journal of Chemical Physics* **2019,** *150* (2), 024502.
25. Harrowell, P., Orientationally ordered glasses via controlled deposition. *Proceedings of the National Academy of Sciences* **2019,** *116* (43), 21341.
26. Lyubimov, I.; Antony, L.; Walters, D. M.; Rodney, D.; Ediger, M. D.; de Pablo, J. J., Orientational anisotropy in simulated vapor-deposited molecular glasses. *The Journal of Chemical Physics* **2015,** *143*, 094502.
27. Bagchi, K.; Jackson, N. E.; Gujral, A.; Huang, C.; Toney, M. F.; Yu, L.; de Pablo, J. J.; Ediger, M. D., Origin of Anisotropic Molecular Packing in Vapor-Deposited Alq3 Glasses. *The Journal of Physical Chemistry Letters* **2019,** *10*, 164-170.
28. Moore, A. R.; Huang, G.; Wolf, S.; Walsh, P. J.; Fakhraai, Z.; Riggleman, R. A., Effects of microstructure formation on the stability of vapor-deposited glasses. *Proceedings of the National Academy of Sciences* **2019,** *116* (13), 5937.
29. Bishop, C.; Thelen, J. L.; Gann, E.; Toney, M. F.; Yu, L.; DeLongchamp, D. M.; Ediger, M. D., Vapor deposition of a nonmesogen prepares highly structured organic glasses. *Proceedings of the National Academy of Sciences* **2019,** *116* (43), 21421.
30. Oosterhout, S. D.; Savikhin, V.; Zhang, J.; Zhang, Y.; Burgers, M. A.; Marder, S. R.; Bazan, G. C.; Toney, M. F., Mixing Behavior in Small Molecule:Fullerene Organic Photovoltaics. *Chemistry of Materials* **2017,** *29* (7), 3062-3069.
31. Ilavsky, J., Nika: software for two-dimensional data reduction. *Journal of Applied Crystallography* **2012,** *45* (2), 324-328.
32. Zhang, F.; Ilavsky, J.; Long, G. G.; Quintana, J. P. G.; Allen, A. J.; Jemian, P. R., Glassy Carbon as an Absolute Intensity Calibration Standard for Small-Angle Scattering. *Metallurgical and Materials Transactions A* **2010,** *41* (5), 1151-1158.
33. Baker, J. L.; Jimison, L. H.; Mannsfeld, S.; Volkman, S.; Yin, S.; Subramanian, V.; Salleo, A.; Alivisatos, A. P.; Toney, M. F., Quantification of Thin Film Crystallographic Orientation Using X-ray Diffraction with an Area Detector. *Langmuir* **2010,** *26*, 9146-9151.
34. Gujral, A.; O'Hara, K. A.; Toney, M. F.; Chabinyc, M. L.; Ediger, M. D., Structural characterization of vapor-deposited glasses of an organic hole transport material with X-ray scattering. *Chemistry of Materials* **2015,** *27*, 3341-3348.
35. Liu, T.; Exarhos, A. L.; Alguire, E. C.; Gao, F.; Salami-Ranjbaran, E.; Cheng, K.; Jia, T.; Subotnik, J. E.; Walsh, P. J.; Kikkawa, J. M.; Fakhraai, Z., Birefringent Stable Glass with Predominantly Isotropic Molecular Orientation. *Physical Review Letters* **2017,** *119*, 095502.
36. Coropceanu, V.; Cornil, J.; da Silva Filho, D. A.; Olivier, Y.; Silbey, R.; Brédas, J.-L., Charge Transport in Organic Semiconductors. *Chemical Reviews* **2007,** *107* (4), 926-952.

37. Shibata, M.; Sakai, Y.; Yokoyama, D., Advantages and disadvantages of vacuum-deposited and spin-coated amorphous organic semiconductor films for organic light-emitting diodes. *Journal of Materials Chemistry C* **2015,** *3* (42), 11178-11191.
38. Yu, L., Surface mobility of molecular glasses and its importance in physical stability. *Advanced Drug Delivery Reviews* **2016,** *100*, 3-9.
39. Samanta, S.; Huang, G.; Gao, G.; Zhang, Y.; Zhang, A.; Wolf, S.; Woods, C. N.; Jin, Y.; Walsh, P. J.; Fakhraai, Z., Exploring the Importance of Surface Diffusion in Stability of Vapor-Deposited Organic Glasses. *The Journal of Physical Chemistry B* **2019,** *123* (18), 4108-4117.
40. Berthier, L.; Charbonneau, P.; Flenner, E.; Zamponi, F., Origin of Ultrastability in Vapor-Deposited Glasses. *Physical Review Letters* **2017,** *119* (18), 188002.
41. De Cupere, V.; Tant, J.; Viville, P.; Lazzaroni, R.; Osikowicz, W.; Salaneck, W. R.; Geerts, Y. H., Effect of Interfaces on the Alignment of a Discotic Liquid−Crystalline Phthalocyanine. *Langmuir* **2006,** *22* (18), 7798-7806.
42. Grelet, E.; Bock, H., Control of the orientation of thin open supported columnar liquid crystal films by the kinetics of growth. *Europhysics Letters (EPL)* **2006,** *73* (5), 712-718.
43. Ruan, S.; Zhang, W.; Sun, Y.; Ediger, M. D.; Yu, L., Surface diffusion and surface crystal growth of tris-naphthyl benzene glasses. *The Journal of Chemical Physics* **2016,** *145* (6), 064503.
44. Mirigian, S.; Schweizer, K. S., Theory of activated glassy relaxation, mobility gradients, surface diffusion, and vitrification in free standing thin films. *The Journal of Chemical Physics* **2015,** *143* (24), 244705.
45. Zhang, Y.; Fakhraai, Z., Decoupling of surface diffusion and relaxation dynamics of molecular glasses. *Proceedings of the National Academy of Sciences* **2017,** *114* (19), 4915.
46. Panova, O.; Ophus, C.; Takacs, C. J.; Bustillo, K. C.; Balhorn, L.; Salleo, A.; Balsara, N.; Minor, A. M., Diffraction imaging of nanocrystalline structures in organic semiconductor molecular thin films. *Nature Materials* **2019,** *18* (8), 860-865.

TOC Graphic

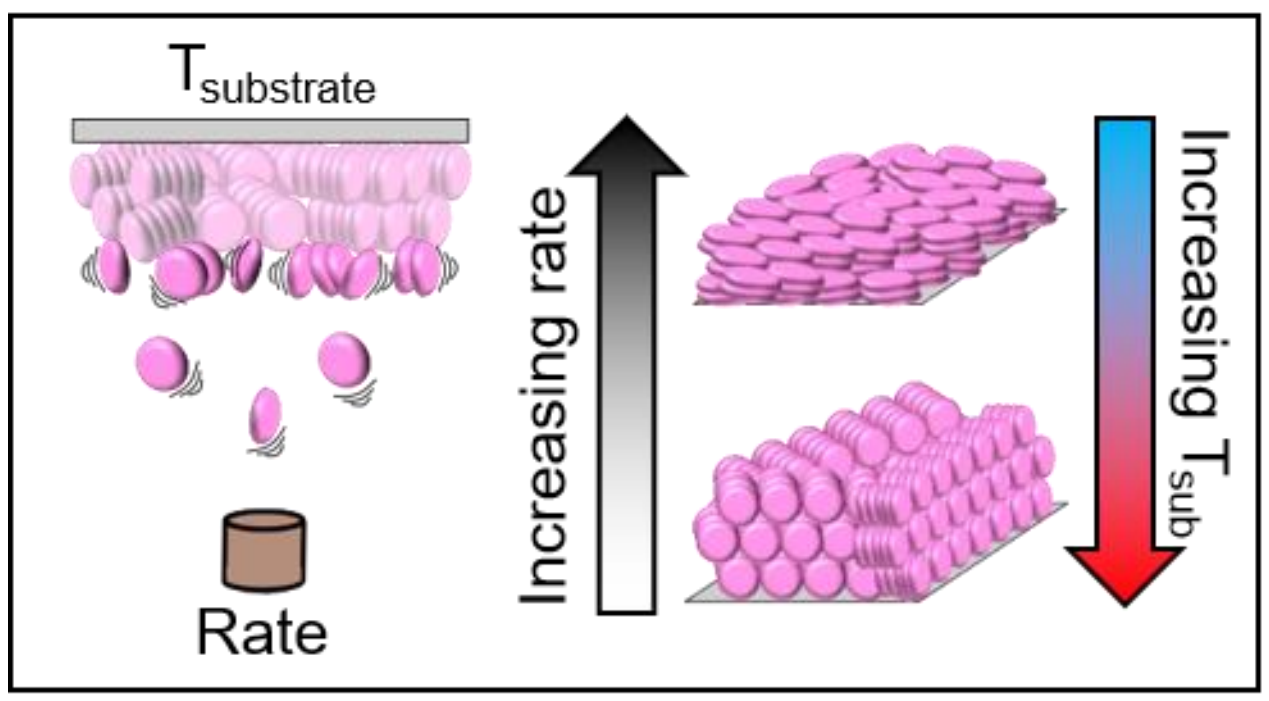

Supplementary Information for "Using deposition rate and substrate temperature to manipulate liquid crystal-like order in a vapor-deposited hexagonal columnar glass"

Authors: Camille Bishop,[1] Zhenxuan Chen,[2] Michael F. Toney,[3] Harald Bock,[4] Lian Yu,[2] M.D. Ediger[1]

Corresponding Author email: camille.bishop@nist.gov

1. University of Wisconsin – Madison, Department of Chemistry, 1101 University Ave, Madison, WI, 53706 (USA)
2. University of Wisconsin – Madison, School of Pharmacy, 777 Highland Ave, Madison, WI, 53705 (USA)
3. University of Colorado Boulder, Department of Chemical and Biological Engineering, Boulder, CO 80309 (USA)
4. Centre de Recherche Paul Pascal, CNRS & Université de Bordeaux, 115, av. Schweitzer, 33600 Pessac (France)

Section 1. $S_{GIWAXS}$ calculation procedures.

To calculate $S_{GIWAXS}$, we first perform a background subtraction. All of the following procedures are shown for X-ray scattering from a glass deposited at 360 K at 0.4 Å $s^{-1}$.

*Background Subtraction*

For all glasses, we find that the majority of the intensity from the q ~ 1.8 Å$^{-1}$ (π-stacking) peak falls between q = 1.65 and 1.85 Å$^{-1}$. We therefore integrate this area (shown in black in Figure S1) at each value of the azimuthal angle $\chi$, with $\chi = 0°$ defined to be along $q_z$. To obtain a low-q background, we perform the same integration on the region from q = 1.55 to 1.65 Å$^{-1}$ (shown in orange on Figure S1). For the high-q background, the same is done for q = 1.85 to 1.95 Å$^{-1}$, shown in red on Figure S1.

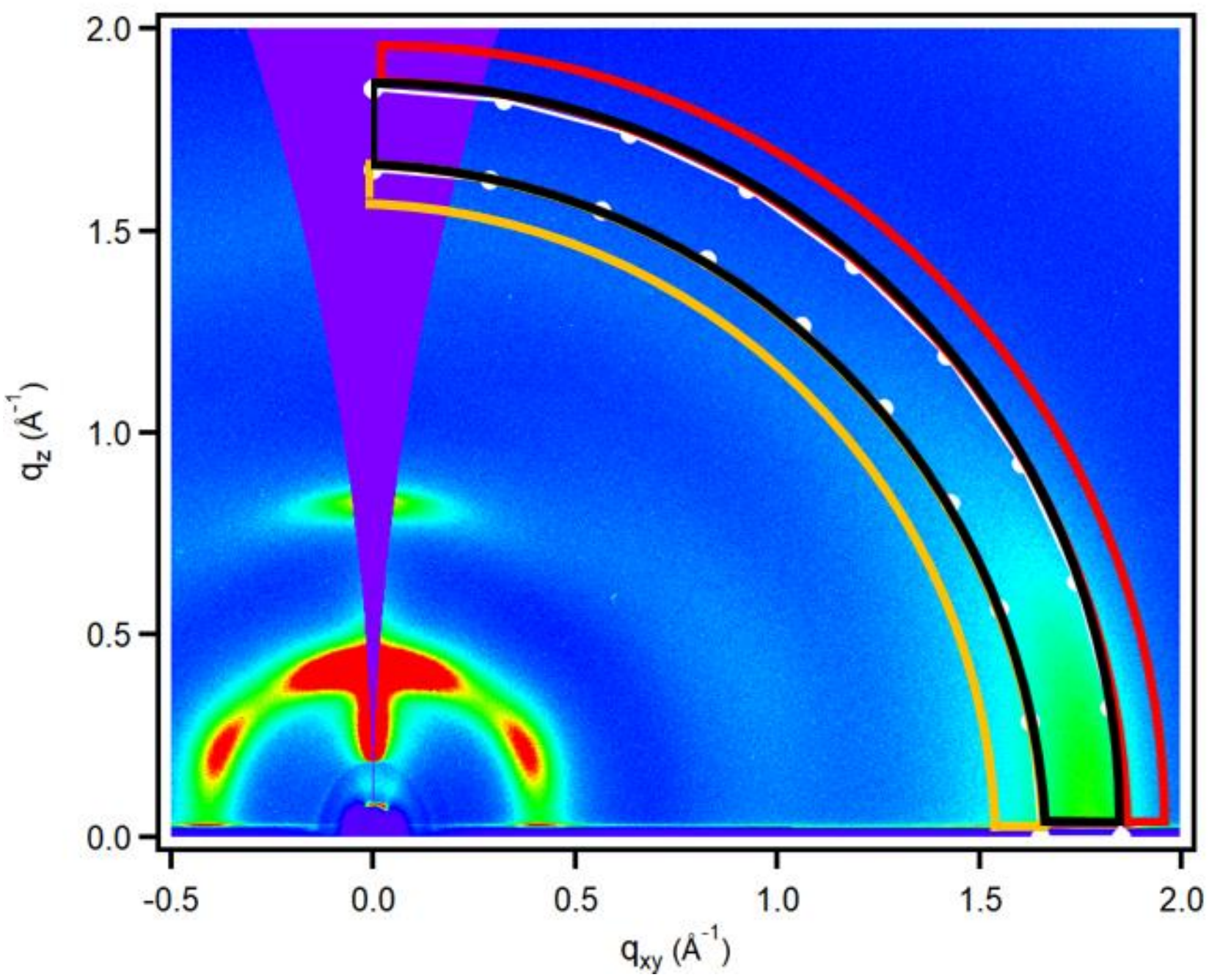

Figure S1. Integrated regions for π-stacking interaction (black), low-q background (orange), and high-q background (red).

These integrations result in the three curves seen in the upper part of Figure S2. To find the background subtracted curve at each value of χ, we subtract the average of the low-q and high-q backgrounds as follows:

$$I(\chi) = I_{q=1.65\ to\ 1.85}(\chi) - \frac{I_{q=1.55\ to\ 1.65}(\chi) + I_{q=1.85\ to\ 1.95}(\chi)}{2}$$

This results in the blue curve at the bottom of Figure S2.

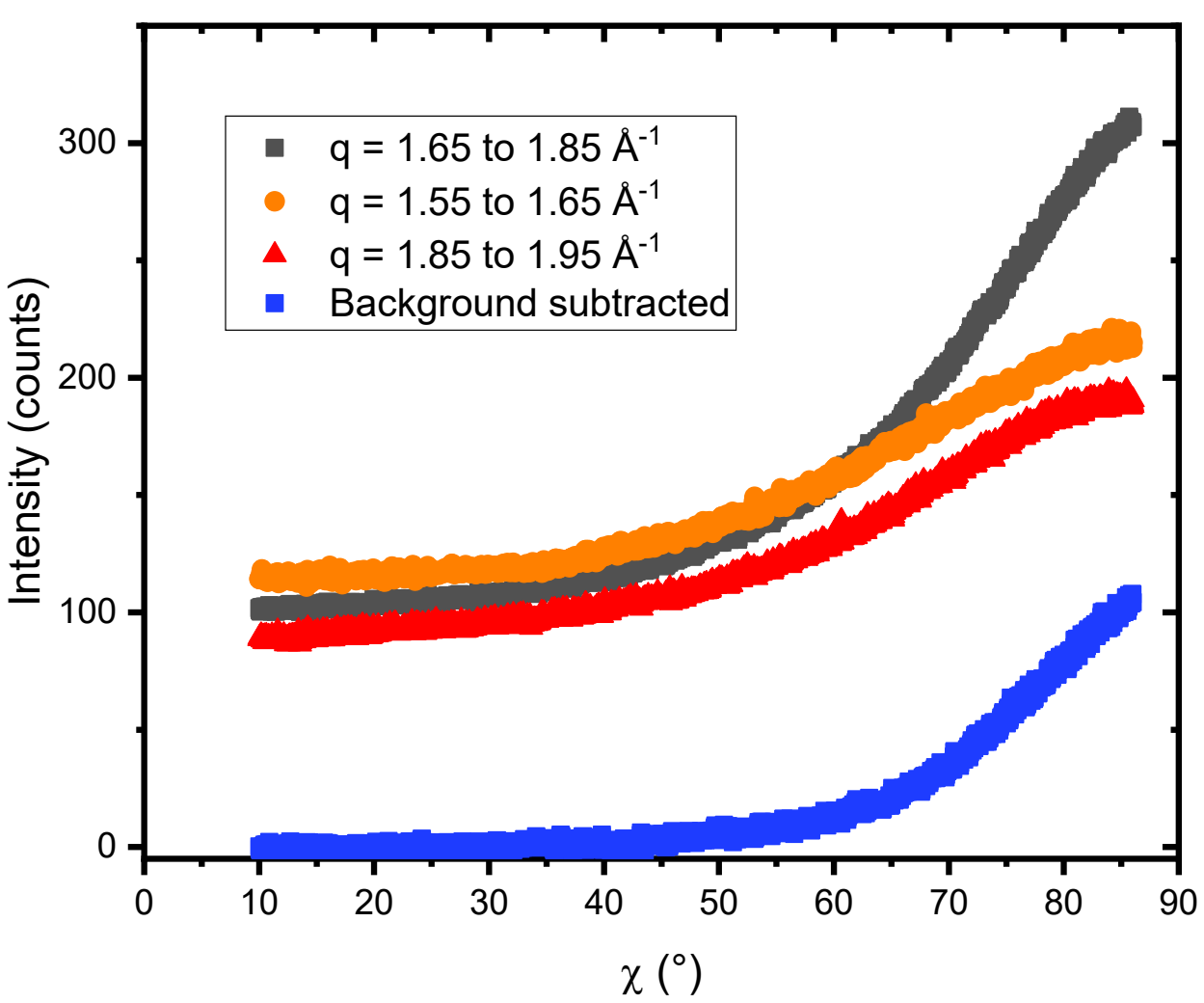

Figure S2. Raw integrations and background subtracted I(χ) curves from Figure S1.

*Empirical extrapolation functions*

Due to the geometry of GIWAXS experiments, the data from χ = 0° to ~10° is inaccessible. Additionally, due to the edge of the sample surface, the data from χ ~ 86° to 90° is dominated by artefacts. For many samples, these regions have sufficient scattering that omission from the calculation of $S_{GIWAXS}$ will result in unphysical values. Therefore, we use empirical functions to extrapolate the scattering into the inaccessible regions, as shown in Figure S3.

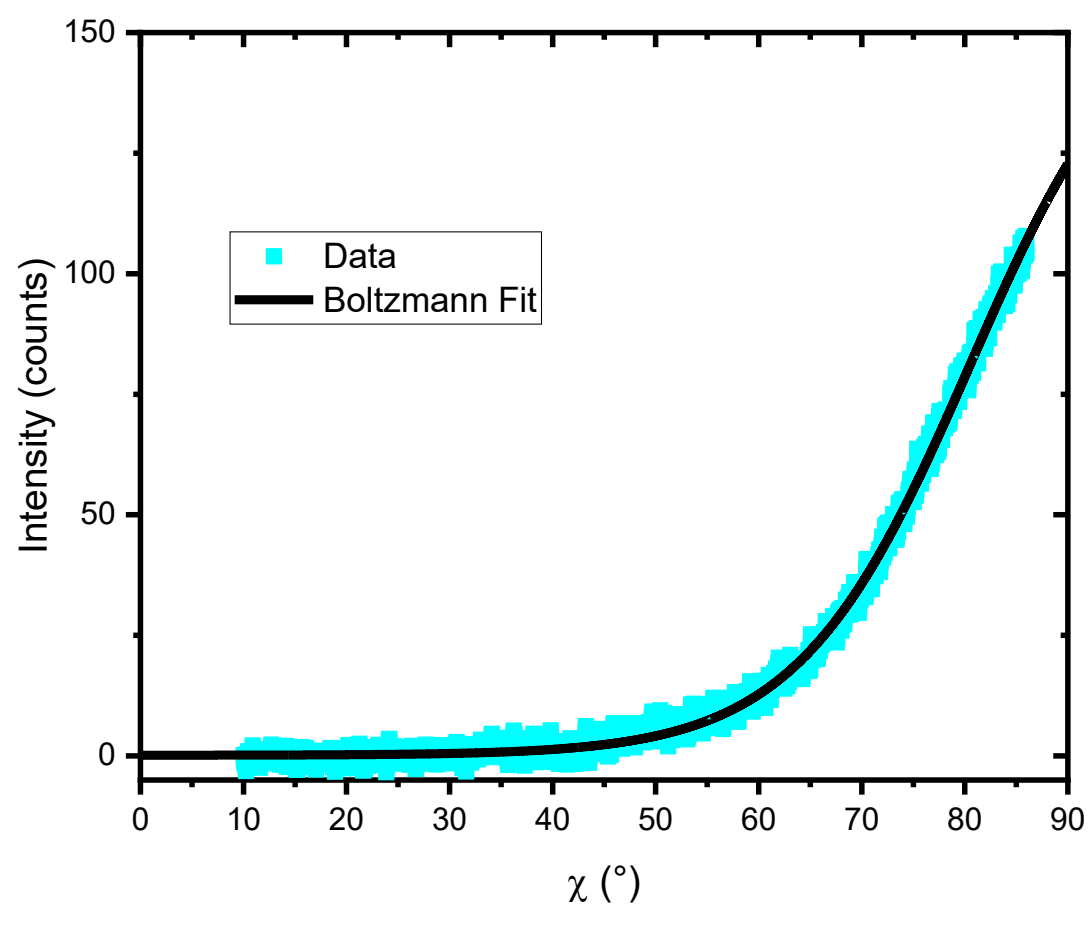

$$I = A + \frac{B-A}{1+e^{(\chi-C)/D}}$$

Figure S3. Background subtracted curve shown in Figure S2, with an empirical Boltzmann fit (equation shown to the right).

In the case of scattering that is extremely focused in-plane, the extrapolations may fail to capture the true scattering. Therefore, caution is required when interpreting results at extreme values of $S_{GIWAXS}$, such as those that do not fall on the linear fit in Figure 2C of the main text.

## Section 2. Determination of $d_{nn}$.

*Selection of azimuthal range.*

To determine the average face-to-face nearest-neighbor distance, we integrate only in the region in which the scattering from this interaction is large. In this way, we find $d_{nn}$ for the dominant structural motif. We apply the background-subtracted curves as shown in SI Section 1 that are used for $S_{GIWAXS}$ determination (which are integrated from q = 1.65 to 1.85 Å$^{-1}$). For each individual sample, we find the maximum scattering intensity. We then find the azimuthal angle χ at which the scattering intensity falls below 50% of the maximum, and define it as $\chi_{min}$, the lower integration limit, shown schematically in Figure S4. For many samples, $\chi_{max}$ occurs at the highest accessible $\chi$, as seen in Figure S4.

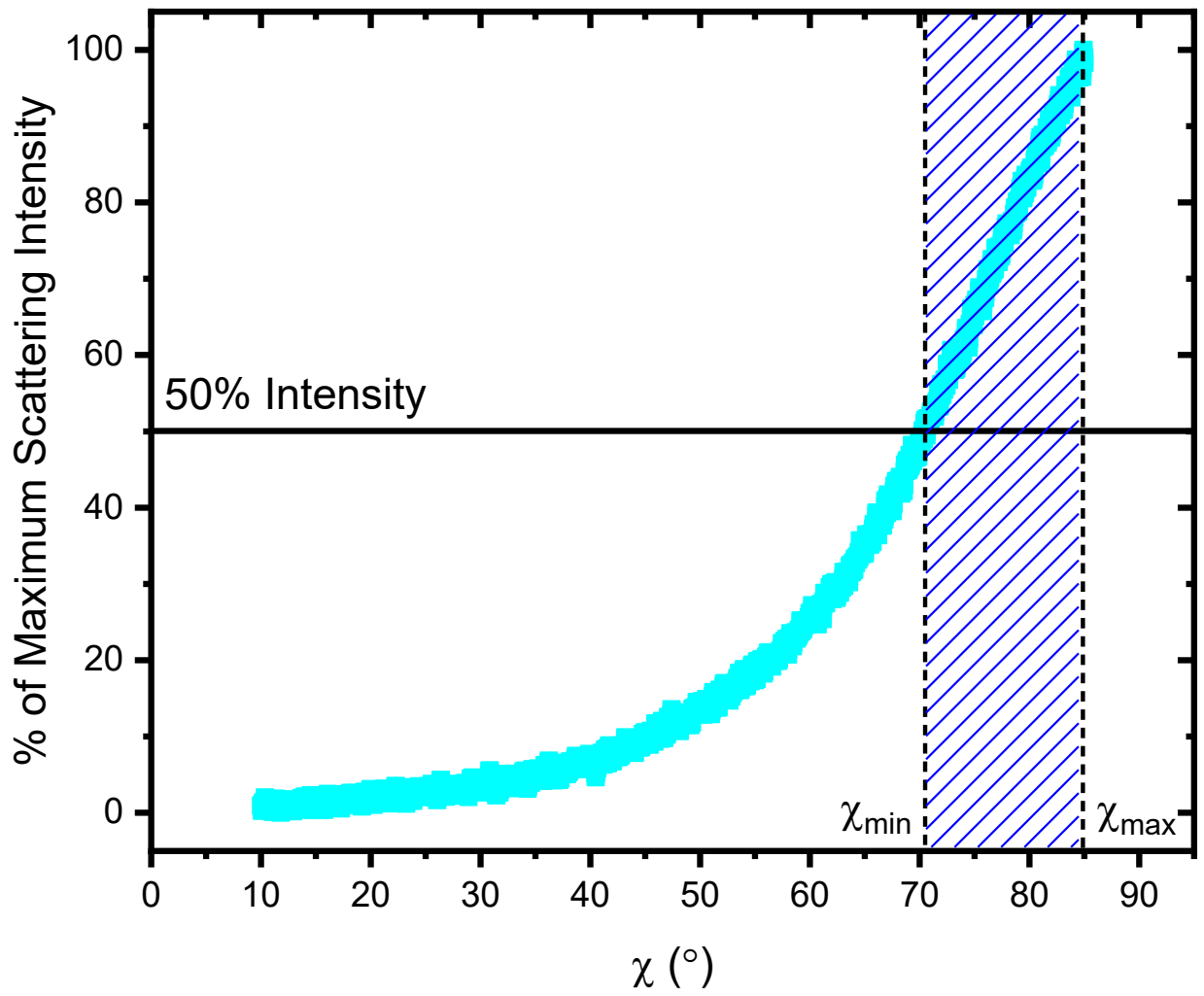

Figure S4. Determination of azimuthal integration limits for $d_{nn}$.

*Determination of peak scattering position.*

Next, the sample is integrated over all $\chi$ at each q in the azimuthal angle range determined by the previous step (in this case, from $\chi = 71°$ to $85°$) to yield the curve seen in the top panel of Figure S5. The curve is smoothed using Savitzky-Golay smoothing with a $q = 0.1$ Å$^{-1}$ window, and then differentiated. We find that the peak position is independent of the smoothing window for smoothing windows of $q = 0.05$ Å$^{-1}$ and higher, as shown in Figure S6. The peak position is determined by the zero of the first derivative:

$$d_{nn} = \frac{2\pi}{q}\bigg|_{\frac{dI}{dq}=0}$$

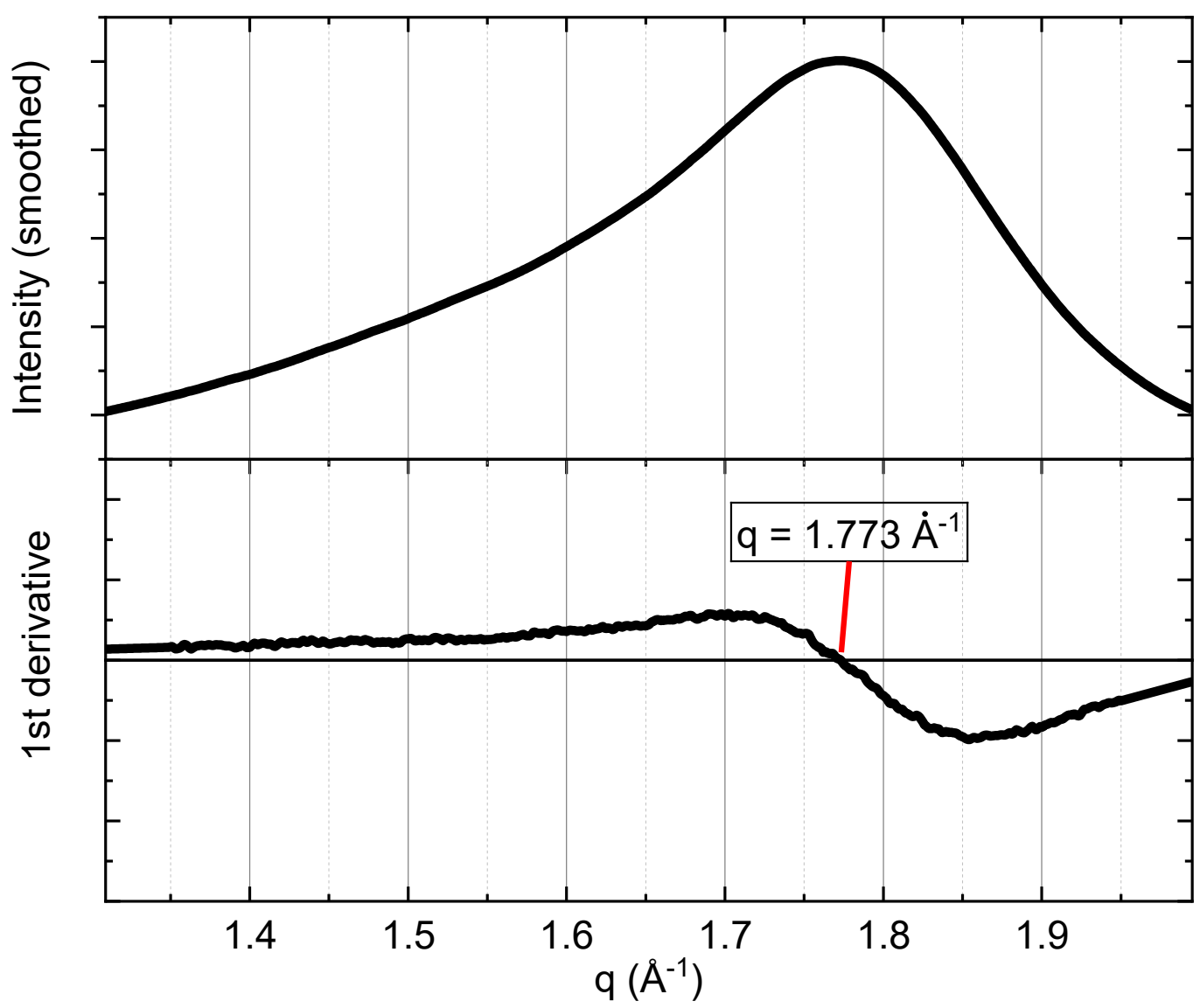

Figure S5. Determination of $d_{\pi-\pi}$ from I vs. q curves.

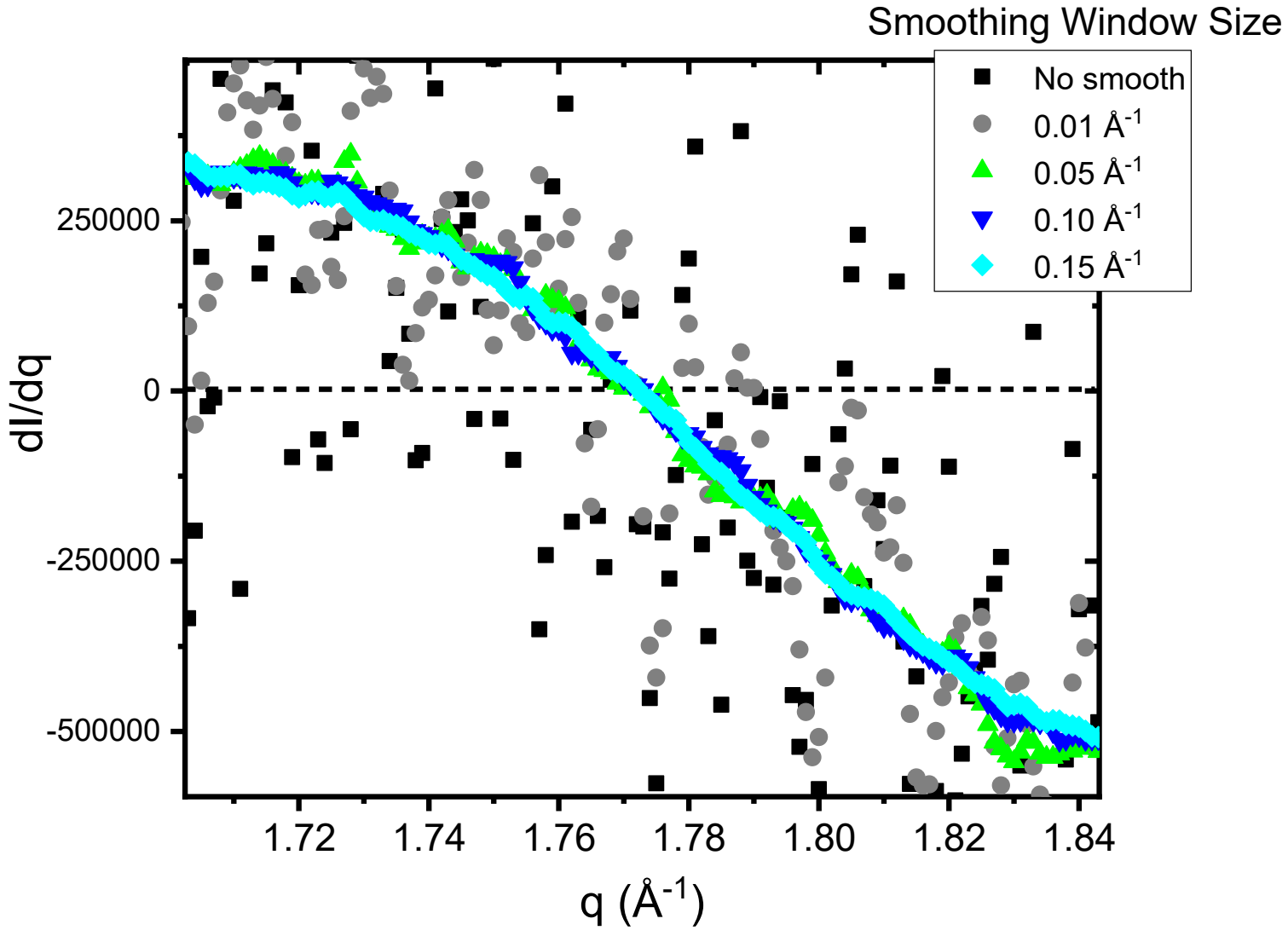

Figure S6. The first derivative used to determine $d_{nn}$ and its effect upon the resulting data. The q-value of the zero of the first derivative, which is the peak position used to determine $d_{nn}$, is independent of smoothing window size above 0.05 $Å^{-1}$.

Section 3. Anisotropy of X-ray scattering and birefringence.

As addressed in the main text, at the deposition condition of $T_{sub}$ = 315 K, deposition rate = 0.8 Å $s^{-1}$, $S_{GIWAXS}$ = 0 while $\Delta n > 0$. The x-ray scattering pattern for this sample is shown in Figure S6. The scattering of the q ~ 1.8 $Å^{-1}$ peak used to calculate $S_{GIWAXS}$ is isotropic, and hence $S_{GIWAXS} = 0$. However, other features in the scattering pattern, such as the q ~ 0.4 $Å^{-1}$ peak are anisotropic. Given that the scattering pattern as a whole is anisotropic, there is no expectation that the birefringence should be precisely zero.

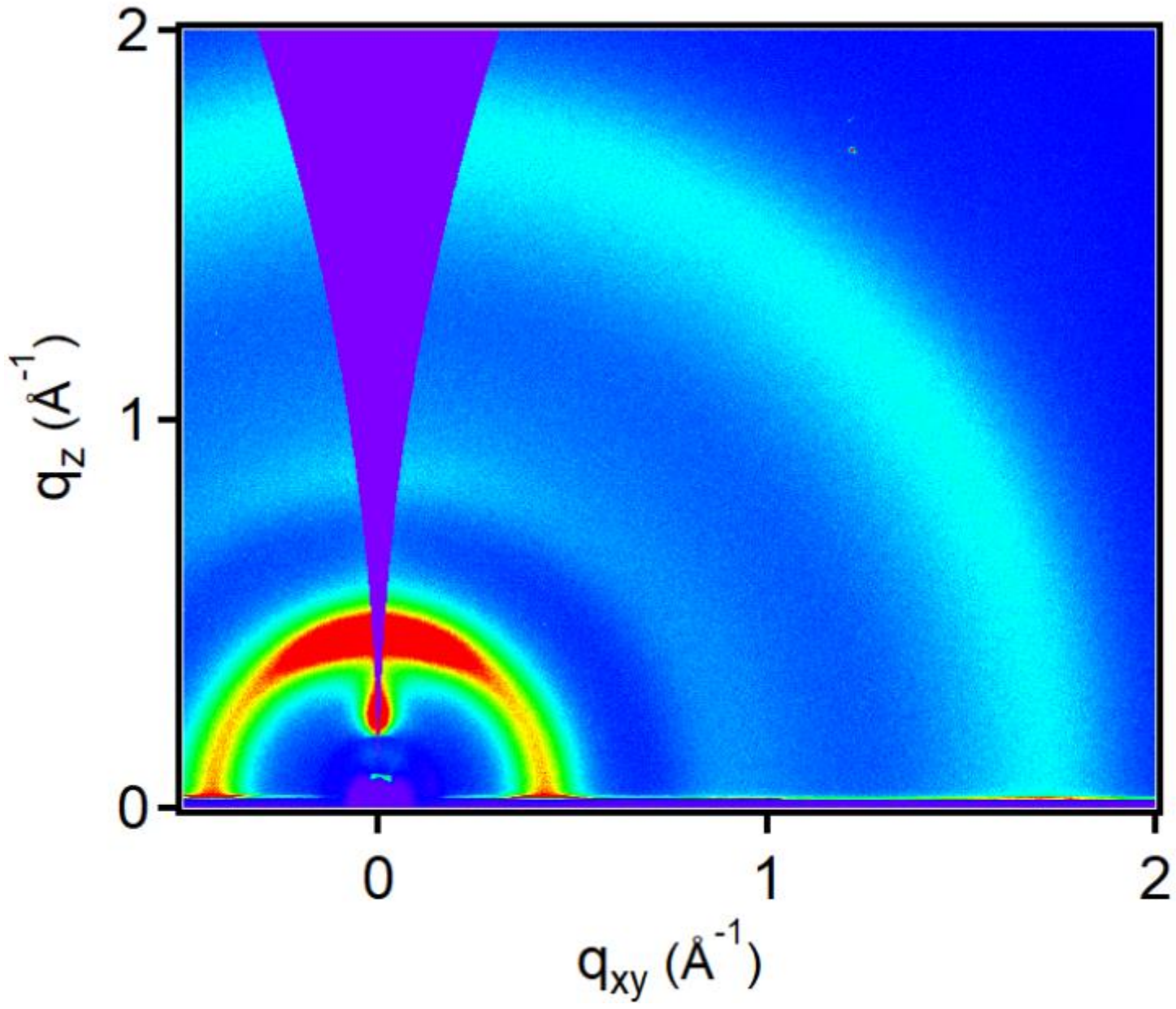

Figure S7. GIWAXS pattern of a glass deposited at 315 K, $10^{-0.1}$ Å $s^{-1}$.

SI Section 4. Edge-to-edge disc spacing of molecules from q ~ 0.4 $Å^{-1}$ peak.

Analogously to the case for face-to-face disc packing seen in Figure 3, we analyzed the spacing for the hexagonal lattice. We first integrate over all $\chi$ = 10° to 85° to plot I vs. q for all accessible scattering space. We then found d for the edge-to-edge hexagonal packing by $d = 2\pi/q_c$, where $q_c$ is the center of a Pseudo-Voigt fit to the peak at q ~ 0.4 $Å^{-1}$ . Results are shown in Figure S8. Figure S9 shows this data plotted against the effective deposition rate calculated using the "shift factor" of 9 K/decade found for hexagonal order in the main text.

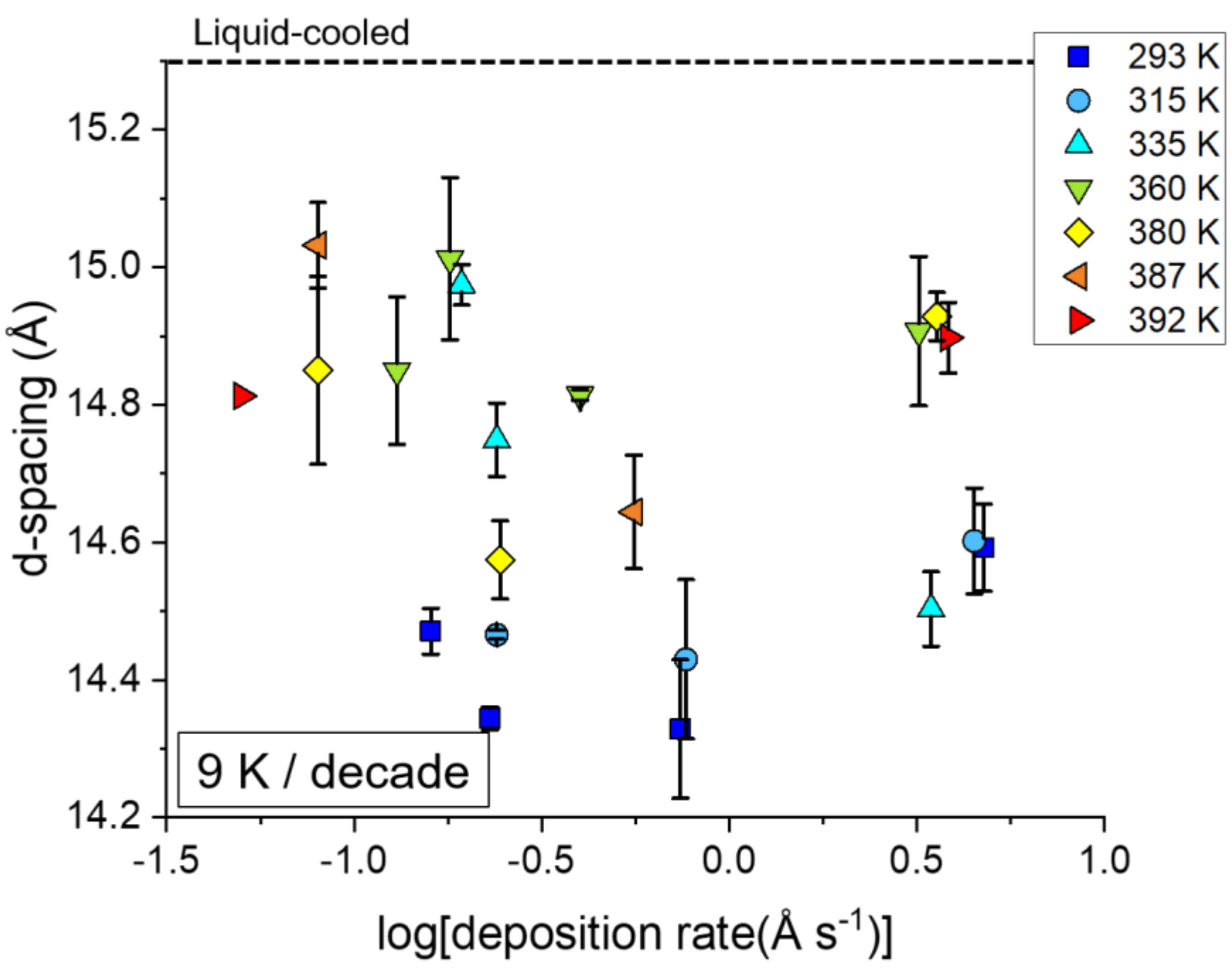

Figure S8. d-spacing of the hexagonal lattice peak found at q ~ 0.4 $Å^{-1}$ as a function of the log of the deposition rate.

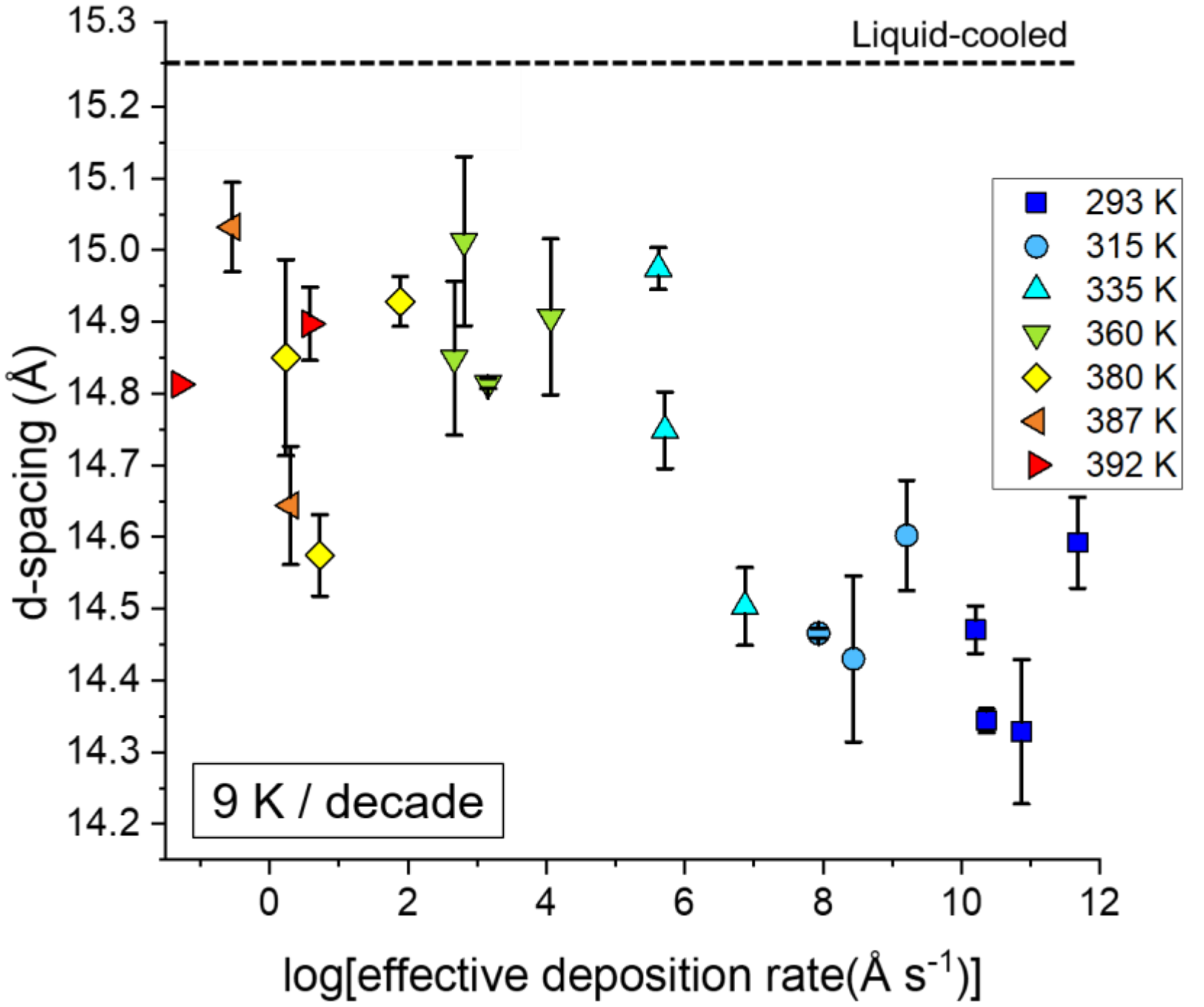

Figure S9. d-spacing of the hexagonal peak as a function of the effective deposition rate.